# In-flight radiometric calibration of the ExoMars TGO Colour and Stereo Surface Imaging System

A. Pommerol[1,*], N. Thomas[1], M. Almeida[1], M. Read[1], P. Becerra[1], C. Cesar[1], A. Valantinas[1], E. Simioni[2], A.S. McEwen[3], J. Perry[3], C. Marriner[4], G. Munaretto[2,5], M. Pajola[2], L. Tornabene[6], D. Mège[7], V. Da Deppo[2], C. Re[2], and G. Cremonese[2].

[1]Physikalisches Institut, University of Bern, Sidlerstrasse, 5, CH-3012 Bern, Switzerland
*To whom correspondence should be addressed: antoine.pommerol@unibe.ch
[2]INAF, Osservatorio Astronomico di Padova, Vicolo dell'Osservatorio, Italy
[3]Lunar and Planetary Laboratory, University of Arizona, 1541 E. University Blvd, Tucson, AZ 85721-063.
[4]School of Physical Sciences, The Open University, Walton Hall, Milton Keynes, UK
[5]Department of Physics and Astronomy "G. Galilei", University of Padova, Padova, Italy
[6]Institute for Earth & Space Exploration, Dept. of Earth Sciences, Western University, London, ON, N6A 5B7 CANADA
[7]Centrum Badań Kosmicznych Polskiej Akademii Nauk, Bartycka 18A, 00-716 Warszawa, Poland

**Abstract**

The Colour and Stereo Surface Science Imaging System (CaSSIS) of the ExoMars Trace Gas Orbiter returns on average twenty images per day of the Martian surface, most of them in 3 or 4 colours and some of them in stereo. CaSSIS uses a push-frame approach to acquire colour images, with four bandpass filters deposited directly above the sensor and an imaging cadence synchronized with the ground track velocity to cover the imaged area with tens of small, partially overlapping images. These "framelets" are later map-projected and mosaicked to build the final image. This approach offers both advantages and challenges in terms of radiometric calibration. While the collection of dark and flatfield frames is considerably enhanced by the frequent and fast acquisition of tens of successive images, mosaics assembled from the adjacent framelets highlight the straylight and changes in the bias of the detector. Both issues have been identified on CaSSIS images, with low intensities overall (up to a few %), but sufficient to generate prominent artefacts on the final assembled colour images. We have therefore developed methods to correct these artefacts that are now included into the radiometric calibration pipeline. We detail here the different steps of the calibration procedure and the generation of the products used for calibration, and discuss the efficacy of the corrections. The relative uncertainties on the bias and flatfield frames are low, of the order of 0.2 and 0.1%, respectively. The uncertainty on the absolute radiometric calibration is of 3%, which is quite low for such an instrument. The straylight adds an estimated ~1% error to the absolute calibration. The residuals after corrections of the straylight and bias offsets are of the order of a few DNs to tens of DNs. As CaSSIS can observe the Martian surface in challenging illumination conditions to provide unique views of the surface at early and late local solar time, residuals from the straylight correction can become noticeable when the absolute signal is very low. As they appear at the level of the noise in very low illumination images, these residuals do not limit the scientific exploitation of the data. For most of the dataset, as the signal in well-exposed images reaches 8,000 DNs in the panchromatic filter and thousands of DNs in the colour filters, the residuals are negligible and CaSSIS provides the best colour images available over many areas covered.

**Keywords**: Mars, Radiometric calibration, Straylight, Pushframe camera, Colour imaging

# 1. Introduction

*The Colour and Stereo Surface Imaging System (CaSSIS)*

The Colour and Stereo Surface Imaging System (CaSSIS, Thomas et al., 2017) on board ESA's ExoMars Trace Gas Orbiter (TGO) has acquired more than 30,000 observations (each with a series of images) of the Martian surface since the end of the aerobraking phase of TGO in April 2018. These images serve the general scientific purposes of the mission and the Exomars program in general and contribute to improving our knowledge of Mars on a variety of topics.

CaSSIS is a moderately high-resolution imaging system, capturing details of the surface at up to 4.5 metres per pixel to produce observations that are typically 9 kilometres wide and 30 to 40 km long (see parameters in Table 1). The instrument uses a 4-mirrors off-axis telescope with a 135mm-diameter primary mirror and effective focal length of 875mm, resulting in a f/6.5 f-number. All four mirrors are silver-coated and powered, in a modified configuration of the classical three mirrors anastigmat (TMA) design to accommodate the allocated volume envelope and reuse of an existing primary mirror. This optical configuration provides low optical distortion (Tulyakov et al., 2018), which is crucial for a stereo imaging system. The structure of the telescope is made of carbon-fiber reinforced polymer (CFRP) to minimize its mass. The entire telescope is wrapped in black multi-layer insulation (MLI) and equipped with a precise temperature control system to maintain the structure and optics at a constant nominal temperature of 293K.

A square 2k CMOS sensor from Raytheon (2048x2048 10µm-pitch pixels) is mounted at the focus of the telescope. It is maintained at a constant temperature of 273K by a large passive radiative cooler and electrical heaters. Mounted on top of the sensor, a glass window with four deposited bandpass filters (Gambicorti et al., 2016) allows the acquisition of images in up to four colours using the "push-frame" imaging approach detailed below. The window is 2mm away and parallel to the detector. Broadband antireflection coating was applied to both the window and the detector to mitigate multiple reflections which could otherwise result in ghost images and crosstalk between filters. To further prevent the latter effect, a black mask was applied on top of the boundaries between the different filters as well as above unused areas of the detector. Theoretical predictions that the antireflection coatings and mask would be efficient at preventing multiple reflection were verified in the laboratory prior to integration of the detector onto the telescope.

The entire telescope together with the sensor and proximity electronics are mounted on a rotation mechanism that can rotate the instrument by 180° in less than 20 seconds. This mechanism serves two purposes. First, it is used to align the sensor with the along track direction. Second, it is used to acquire nearly simultaneous observations of the surface. For that purpose, the line of sight of the telescope is tilted by 10° relative to the rotation axis. In this configuration, CaSSIS takes a first image of the scene in the forward direction, then rotates by 180° degrees and takes a backward image of the same scene about 45 seconds later. This results in an ideal stereographic convergence angle of 20°.

CaSSIS acquires an average of 20 observations per day, representing about 40 Gbits of data. However, the number of images taken varies strongly depending on the distance from Earth to Mars, which affects the downlink capability of the spacecraft. Also, we may choose to limit imaging when or where dust storms prevent useful imaging of the Martian surface.

Targets are carefully selected by the CaSSIS science team based on the scientific interest of the area and the observation conditions available to CaSSIS. Depending on the properties of the target and the scientific objectives, a variety of "imaging modes" is available for the planner. An imaging mode is the combination of parameters that defines a particular observation, these include the number of colour channels (up to four), image compression settings, and final swath size. In addition, CaSSIS is capable of acquiring nearly-simultaneous stereo images that can be combined to produce Digital Terrain Models of the surface and anaglyphs (Simioni et al. 2021, Re et al. this issue).

*Imaging approach and nomenclature*

Common to all possible imaging modes is the general scheme for colour imaging with CaSSIS which combines a "push-frame" technique with colour filters deposited on a glass window fixed in front of the detector (Figure 1). In this approach, the detector acquires images repetitively, with a repetition time synchronized to the ground-track velocity and range to the surface so that adjacent images can be mosaicked. The repetition time (200 to 300 ms) is calculated in order to keep an overlap (10 to 15%) between successive acquisitions, which is later used to perform a bundle adjustment to optimize the mosaicking and mitigate the effects of imprecise timing, misalignment, or jitter (Perry et al., this issue). At the acquisition frequency needed, the entire detector cannot be read fast enough, therefore, up to six predefined Regions Of Interest (ROIs) are read instead, and these data are transmitted. The four colour filters are deposited on a fused silica glass window that is fixed 2 mm away from the detector, and a predefined ROI is located behind each of the four filters. An image acquired in one of these four regions of interest is called a "framelet". Depending on the imaging mode selected by the planner, one to four framelets can be acquired simultaneously within an "exposure" of the sensor (an electronic global shutter is used so that all pixels are physically exposed to light at the exact same time and the exact same duration). Between 20 and 50 (but frequently 30 to 40) exposures are acquired, which together define an imaging "sequence". A sequence therefore lasts about 10 seconds depending on the exact number of framelets.

The resulting sets of framelets are bundle adjusted to insure good registration of colour images and then map-projected and mosaicked independently for each filter. This results in one to four single-filter images that can be combined to produce the final colour images. A typical CaSSIS colour image is produced from 30 to 40 acquisitions with 2 to 4 colour filters. In rare cases, monochromatic images with the PAN filter only are taken. In addition, in imaging modes that require all four filters to be acquired simultaneously, it is necessary to reduce the width of the framelets because of internal bottlenecks in data transmission within the instrument. Depending on the altitude of the spacecraft at the time of acquisition, the primary control on the required acquisition frequency, this restriction can also apply to the simultaneous acquisition of three framelets. In the Southern Hemisphere, where the altitude is lower and the ground track velocity is higher, an all 4 filter image is typically required to be 1344 pixels in width. In the mid- to high northern latitudes, it is possible to achieve 3 filters images with the full 2048 pixels width. The push-frame approach used by CaSSIS for colour imaging is illustrated in Figure 1. Pictures of the detector and filters assembly are shown in Thomas et al. (2017).

CaSSIS can compress sets of images buffered before transmitting them to the spacecraft, using either lossless or lossy compression. Lossless compression results in a compression ratio of 1.75 while the compression ratio with lossy compression is higher, depending on chosen settings. A compression ratio of 3 is the most frequently used. This is used to optimize the number of images taken depending on the downlink capability of the spacecraft.

It is also possible to digitally bin the pixels in the instrument in order to reduce the size of the data, but this reduces the spatial resolution of the images. A 2x2 pixel binning reduces the resolution by a factor of two and the image data size by a factor of 4. Although this possibility has been occasionally used at the beginning of the primary science phase, it was found to produce significant calibration issues caused by bugs in the flight software. In addition, it became clear that the lossy compression was more efficient (less lossy) than binning to reduce the size of the data and the use of binning was largely suspended. Later in the mission, after the upload of an updated version of the flight software that fixed the binning issue (among other problems), binning was used to acquire targets in low-light conditions. In these conditions, the integration time is usually limited by the 1-pixel smearing time, which is about 1.5 ms. But using 2x2 binning, one can double the integration time (up to ~3 ms) to collect more photons, and thereby increase the signal-to-noise (SNR) of the images.

The SIMBIO-SYS instrument (Cremonese et al., 2020) developed for ESA's BepiColombo mission to Mercury uses a "push-frame" approach with colour filters deposited directly above the detector. Since CaSSIS uses a flight-spare of a SIMBIO-SYS sensor, it adopted this approach from it. While this approach is relatively recent for planetary missions, CaSSIS is not the first instrument to use it. The Thermal Emission Imaging System (THEMIS; Christensen et al., 2004) of NASA's 2001 Mars Odyssey was the first instrument to image Mars at visible and infrared wavelengths using this technique. The visible channel THEMIS-VIS shares many characteristics with CaSSIS. The spatial resolution of CaSSIS is about 3 times better than THEMIS-VIS (18m/pixel). THEMIS-VIS has one additional colour band (5 bands from 0.425 to 0.86 µm) compared to CaSSIS, but CaSSIS has a wider sensitivity range spanning 400-1000nm. The Wide-Angle Camera of the Lunar Reconnaissance Orbiter (Robinson et al., 2010) is another example of a recent instrument using the push-frame approach for colour imaging. In fact, this approach is more often chosen with wide-angle imagers – e.g. the MARCI camera on NASA's Mars Reconnaissance Orbiter (MRO) or the JunoCam on NASA's Juno spacecraft – while for most high-resolution narrow-angle imagers, the push-broom method remains the most popular approach. Some of the particular challenges associated with push-frame cameras have been noted and discussed by Anderson and Robinson (2009).

*Laboratory calibration efforts*

Despite a tight development schedule, CaSSIS was subjected to a series of tests and ground calibration procedures following assembly and prior to integration onto the spacecraft. The methods and results were detailed by Roloff et al. (2017), and some of the main results and their implications are summarized later in this section. As the approach used by CaSSIS for colour imaging is similar to that used by the High-Resolution Imager (HRIC) and the stereo imager STC of BepiColombo's SIMBIO-SYS, many of the calibration results obtained with the SIMBIO-SYS detectors in the laboratory (Slemer et al., 2019), and in flight (Zusi et al., 2019) are also applicable to CaSSIS. In particular, it was noted on the SIMBIO-SYS detectors that the bias of the detectors can vary significantly during a series of acquisitions, depending both on the exposure time and the acquisition frequency, an effect related to the reset capacity of the detector. Particular attention was therefore placed on this potential issue during the design, construction and testing of CaSSIS.

The measurements of bias and darks with the CaSSIS detector under various conditions showed that the amount of thermally generated dark current is negligible at the 273K operational temperature of the CaSSIS detector and the exposure time used for Mars imaging (<1.5ms). These tests also revealed that the CaSSIS detector suffers from the same issue already identified on the SIMBIO-SYS detectors. When acquiring a series of bias or dark measurements, the signal read decreases with the first exposures before it stabilises. The number of exposures before stabilisation as well as the amplitude of the excess signal measured in the first exposures depend on both the frequency of the acquisitions and the exposure time. Fortunately, the issue is not severe for CaSSIS images of Mars as the maximum excess signal is only one to two DN at the maximum exposure time of 1.5ms. Nevertheless, this effect must be taken into account when imaging other objects with longer exposure times, such as reference stars for absolute calibration purposes (Thomas et al., 2022).

Flatfields were also measured in the laboratory using a large integrating sphere, which revealed the presence of shadows from dust particles on the optics of the telescope, reducing locally the sensitivity by up to 10%. The linearity of the detector was verified and found to be excellent (Pearson product-moment correlation coefficient PCC>0.99). A limited number of defective pixels was identified and used to refine the position of the windows read below the colour filters.

The relative spectral response of the instrument was characterized within each of the four colour filters, and shown to be consistent within 5% with a model of the instrument spectral response built from information provided by the manufacturers of the different components (Roloff et al., 2017; see also Fig. 1B here). A few attempts at characterizing the absolute response of the instrument or the filter-to-filter relative spectral response were unsuccessful for various technical reasons and the lack of time prevented us from obtaining robust results on these aspects.

*In flight calibration campaigns*

In-flight testing and calibration campaigns were executed four times before the primary science mission: (1) shortly after launch ("commissioning", COM, April 2016) (2) around the half-time of the cruise phase ("mid-cruise checkout", MCC, June 2016), and twice following Mars orbital insertion on the initial parking orbit ("Mars capture orbit", MCO); (3) first on two consecutive orbits (#9 and # 10) in November 2016, and (4) on a single orbit (#50) in March 2017. During these calibration campaigns, CaSSIS acquired a variety of images of dense starfields, reference stars with different colours and brightnesses, a series of images of Phobos crossing the detector in both directions from MCO orbit # 10 and images of Mars with variable resolution from the initial elliptical parking orbit. In addition, a campaign to point CaSSIS at Jupiter for calibration purposes was implemented and executed in December 2019 during the primary science phase, and two additional reference stars were observed in January 2021.

Well-exposed images of individual bright stars, some of them spectrophotometric standards ($\pi^2$ Orionis, $\chi i^2$ Ceti) were used to measure the Point Spread Function (PSF) of CaSSIS ([Fig. 2](#)), showing an average Full Width at Half Maximum (FWHM) of 1.2 to 1.3 pixels in the PAN, RED and NIR filters and 1.4 to 1.5 pixels in the BLU filter ([Gambicorti et al., 2017](#)). The very first series of measurements after launch (April 2016) also demonstrated the efficiency of the temperature regulation of the telescope. Imaging started with a cold telescope, showing a larger PSF with an average FWHM between 1.5 and 2 pixels. As heaters warmed up the telescope to its operational temperature of 293K, the FWHM of the PSF progressively decreased to the smaller values mentioned above ([Fig. 2](#)). The observed change in focus is the result of the thermal expansion of the carbon fiber structure of the telescope. The images of photometric reference stars and Jupiter were also analysed to refine the absolute photometric calibration of the instruments ([Thomas et al., 2022](#)).

Starfield images were used to derive the precise boresight vectors for the centres of all four colour windows as well as the geometric distortion map of the field-of-view. Both sets of data are crucial for successful mosaicking of the framelets, final assembly of the colour products, and stereo reconstruction. This part of the calibration was described by [Tulyakov et al. (2018)](#) and the results are part of the SPICE instrument kernel of CaSSIS, archived and distributed through ESA's Planetary Science Archive (PSA) and NASA's Planetary Data System (PDS). The geometric distortion of the instrument was found to be very close to that expected from the optical simulations of the telescope manufacturer. These assessments resulted in the refinements of the focal length of the telescope to a value of 875±2 mm.

## 2. Formats and Methods

Figure 3 provides a schematic overview of the entire calibration process and associated data levels. All CaSSIS data decoded from the raw spacecraft telemetry are stored in a format compliant with the Planetary Data System (PDS) version 4 format (Planetary Data System, 2016; Besse et al., 2018). Each individual framelet is stored in a binary data format (.dat file extension) associated to a PDS v.4 XML header (.xml file extension) which contains both the information required to read the data file and the header or auxiliary information on the image acquisition, including instrument settings, and geometric and geographic information about the imaged area. Information for the latter is derived from the instrument operation timing using the SPICE toolkit developed by NASA's Navigation and Ancillary Information Facility (NAIF).

The filenames are in the form of CAS-M02-2018-05-30T20.59.49.711-BLU-XXYYY-ZZ where M02 indicates that the image was acquired during Mid-Term Planning (MTP) period #2, followed by a date and time stamp for the acquisition, the filter window (BLU, PAN, RED or NIR), a five-digit integer -XXYYY- and a final two-digit integer –ZZ (XX is a counter for the spectral window used starting with 0. YYY is a counter for the framelet position within the image also starting with 0, and ZZ is the calibration level).

The first level of data, level 0, corresponds to the raw data decoded from the telemetry and converted to the PDS format without any operations. These files are stored on a server at the University of Bern, sorted by MTP (Mid-Term Planning, one month of data) number, STP (Short-term Planning, one week of data) number, and finally a "boot number", which typically ranges from one to eight and corresponds to a daily reboot of the instrument planned to mitigate sporadic issues that could result in data loss. A typical boot folder contains about 20 CaSSIS observations. With an average of 30 framelets per image, per colour, and an average of 3 colours per observations, a typical boot folder contains 1800 framelet data files and just as many headers. Note that all issues that made daily boots of the instruments necessary have been solved by a software update in July 2020 and the frequency of reboots of the instrument has been regularly decreased, down to a value of one reboot per week since April 2021. For this reason and in order to keep the number of files in folders manageable, "period folders" have now replaced "boot folders".

The second level of data, level 1, corresponds to calibrated products after the bias subtraction, flatfield division, conversion of the signal in Digital Numbers (DN) to physical units (radiance factor, I/F), and interpolation of bad pixels. Both the bias and flatfield frames are derived from the in-flight observations detailed later in Sections 3.1 and 3.2 of this manuscript, respectively.

The third level of data, level 1c, corresponds to the calibrated level 1 products with additional corrections for straylight and framelet-to-framelet offsets. This level is the final stage of the radiometric correction, and level 1c data are then directly ingested into the geometric calibration pipeline that corrects the geometric distortion (Tulyakov et al., 2018), mosaics the individual framelets and generates the colour products (Perry et al., this issue).

In order to process the large and numerous CaSSIS dataset efficiently, filter the suitable data, produce statistics and generate calibration products, a computing server with the file server mounted is used. A series of programs written in the Interactive Data Language (IDL) are used to manipulate the database, read the files, and generate both products and reports. Central to these tasks is a structured mapping of all files grouped by observations (all files which belong to a given CaSSIS colour image – identified by a unique "CTF_ID" identifier) and all information on the observation that is relevant for calibration. Optimization of the routines and the structured database make the handling of the large dataset very efficient. All calibration and analysis tasks are run on the entire dataset (more than 13'000 individual observations at the date of the latest processing for the writing of this paper) within a few hours. The IDL programs used to generate calibration products are kept under strict configuration control using the git version-control software. In addition to calibration products in binary form, the IDL programs also generate reports as comma separated value (csv) tables that list all observations used and other relevant information. Experimental versions of the radiometric and geometric calibration pipelines are also installed on this server for the purpose of testing the effects of new and different calibration products and procedures without having to interfere with the currently operational version of the calibration pipeline.

## 3. Steps of the radiometric calibration procedure

### 3.1. Bias subtraction

It was shown during the ground calibration campaign that at the nominal operational detector temperature of 273 K, the level of thermally-generated dark current was negligible for the exposure times relevant to the observations of Mars (Roloff et al., 2017). A constant bias (or offset) frame must therefore be subtracted from every acquisition, independent of the exposure time and exact temperature of the detector. This bias frame corresponds to the fixed-pattern generated by the electronics of the sensor itself. The absence of any thermally-generated dark current when imaging Mars at a nominal exposure time <1.5 ms was verified again during the primary science phase, when acquiring images of the night-side of Mars with slightly different integration times to generate the in-flight bias products.

On the ground, the bias frame was acquired while CaSSIS was operating inside a large thermo-vacuum chamber at low temperature with the detector regulated at nominal temperature. The lab was kept dark and all optical interfaces of the chamber carefully masked. In flight, new bias frames were produced by averaging CaSSIS images acquired over the night side of Mars. Because of the non-Sun-synchronous orbit of TGO, it is important for this purpose to only select observations acquired at phase angles larger than ~120°, to avoid contamination of night-time data by scattered light from the atmosphere close to the terminator (90°). Bias frames have been acquired on the ground prior to flight, during in-flight commissioning and calibration, and at different times of the mission primary phase, in order to ensure that the bias frame was not variable with time. Night-time images specifically planned for this purpose are acquired with two different configurations: (1) two spectral windows per acquisition and nominal cadence, or (2) all four spectral windows at a time but reduced cadence. In both cases, images are acquired with full detector width (2048px) and either no or lossless compression. In addition, bias frames are acquired with and without pixel binning (2x2), and temperature sensors continuously monitor the temperatures of the detector itself, and of its proximity electronics, which are susceptible to affecting the dark level of the detector.

The bias frame (2019-03-08) used for the latest reprocessing of the dataset performed in the first half of 2022 was obtained by averaging the five images with the lowest median signal within the entire dataset, for each filter (Table 2). These night-side images were captured within two consecutive weeks (stp019 and stp020, in Aug.-Sep. 2018). Because each of these images is composed of 30 framelets, the bias frame is averaged from 150 individual detector acquisitions, resulting in a good signal-to-noise ratio. The bias frame shows deep furrows in the y-direction of the detector (Fig. 4) with amplitudes of up to 500 DN. When averaged along the x- and y- directions, the bias signal shows gradients in both directions (Fig. 4).

In December 2019, a new bias product (2019-12-20) was generated using additional data taken at different times of the mission. The selection process was also changed from a fixed number of images with the lowest signal to a variable number of images within a permitted range of DN above the lowest median signal measured to date. This threshold was fixed at 12 DN, which resulted in the selection of 11 images in PAN, 8 in RED, 7 in NIR and 13 in BLU.

The comparison between the bias frame currently in use (2019-03-08) and the newest product (2019-12-20) did not reveal any significant differences between the two bias frames. The shallow gradient in the y-direction of the detector is only slightly different, with amplitudes of up to 0.5 DN / 100 pixels (Fig. 5). A difference of 0.5 DN is also visible within the oblique line defect that crosses the detector below the PAN filter. Because a larger number of images were averaged with a larger range in the median signal DN, the overall median value of the bias frame is also larger by 2 to 7 DN (depending on the filter). Tests in which we applied a different version of the bias frame to calibrate a CaSSIS image with a very low signal did not reveal any difference in the quality of the processed images. An update of the bias frame was therefore not deemed necessary.

Dark frames were also similarly acquired, but performing a 2x2 binning directly in the proximity electronics of the detector. The comparison of these frames with binning applied on the ground, to the full-resolution bias frame did not reveal any significant differences.

The dark current rate was verified from the images of dense starfields acquired during near-Earth commissioning and mid-cruise checkout to characterize the geometric distortion (see Tulyakov et al., 2018). At the long exposure time (1.92s) of these images, the background level of ~7000 DN in the first acquired framelet is consistent with the dark current rate (600 DN/s) and the increased bias (5680 DN) measured in the laboratory by Roloff et al. (2017). The dark current rate was not characterized further however, as for all calibration products we systematically proceed by subtracting from the image of the object (star, planet, moon) a background image taken in the exact same conditions (integration time, position in the sequence of images acquired) by moving either the telescope or the spacecraft in between the acquisitions. We find this procedure more precise and reliable than having to rely on a dark current rate. See Thomas et al. (2022) for additional details.

The overall error on the bias frame (on average over the entire detector) is estimated from the standard variation of the mean bias value calculated from all observations collected over the night side of Mars (defined using a conservative value of 120° of phase angle to avoid limb illumination close to the terminator). We obtain a value of standard deviation of ~7 DN, which represents a relative error of ~0.2% of the bias level. This is also the order of magnitude of the differences observed between the current and the newest product (Fig. 5). The pixel-to-pixel variability is at least one order of magnitude smaller.

*3.2. Flatfield division*

In order to correct for the pixel-to-pixel variability of the sensitivity of the detector, as well as for fixed-patterns caused by the optics, all images are divided by a flatfield frame immediately after the subtraction of the bias frame. The flatfield frame is obtained by averaging as many images of the Martian surface as possible to blur the signal from the surface and retrieve the fixed-pattern intrinsic to the instrument. This process takes particular advantage of the push-frame approach used by CaSSIS, as an average of 30 to 40 framelets is available for each single observation, which allows significant statistics to be obtained quickly. In homogeneous scenes, the flatfield pattern is apparent and recognizable on an average of all framelets performed on a single observation, which allows us to accurately track temporal changes in the flatfield.

At this stage, filtering of the dataset to select all suitable images and reject unwanted images is crucial. Unwanted images include images affected by saturation, even of a very limited number of pixels, or scenes that present very strong contrasts, such as brightly illuminated slopes next to shadowed regions. Images showing scenes with very low contrast such as homogenous terrains or images affected by dust and clouds often make excellent contributions to the flatfield. Images collected during the Planet-Encircling Dust Event (PEDE) of 2018 (shortly after the beginning of the primary science phase) were particularly useful to improve the flatfield operation early in the mission.

The production of the initial flatfields used during the first months of the mission was largely a manual process, where all averaged observations (with all the individual framelets from an observation stacked) would be analysed by eye for homogeneity and saturation. As the number of observations in the database grew steadily, this manual filtering of the data became unsustainable and an automatic procedure was developed, mimicking the manual procedure performed initially. After averaging all framelets for a given image, we calculate and analyse the average vertical and horizontal brightness profiles through the resulting averaged image. The standard deviations of these profiles are then used to select or reject the corresponding image. The thresholds for selection and rejection are then based on visual inspection of multiple images, and testing and comparison of the products resulting from the use of different values.

The entire process is schematized in Figure 6. Table 3 shows the number of images used for producing the flatfield frame currently in use, which is shown in figure 7 and other versions of the flatfield.

The flatfield shows a distinctive checkboard pattern with squares ~20 pixels wide. This pattern is more visible at the bottom and top of the detector, through the PAN and BLU filters respectively, than in RED and NIR. A linear "scratch" is visible in the lower right corner of the PAN filter area. The most enigmatic feature is located close to the centre of the NIR filter and consists of a bright halo surrounded by a closed dark line displaying straight and curved sections. All these features seem to be stable and have not varied with time, including the characterizations performed in the laboratory with an integrating sphere prior to flight (Rollof et al., 2017).

About 20 round features with typical diameters of 10 to 30 pixels are also seen at different positions across the detector. The reduction in sensitivity over these spots is variable and of the order of a few percent, with a maximum of about 10% for the small round feature in the upper right corner of PAN. Most of these features moved, appeared or disappeared between the last flatfield measured in the laboratory and the first flatfield derived from flight data. They are likely shadows from dust particles within the optics of the telescope that move with vibrations and shocks.

The production of different versions of the flatfield over the course of the mission using an ever-increasing number of images reveals the stability of the flatfield. The only significant change detected so far is the appearance of two new dark spots at the bottom of the RED filter between stp034 and stp036 (Fig. 7), again likely to be produced by a single dust particle. The impact and reduction of the sensitivity within these spots is very small, of the order of 0.5 and 0.3% respectively so that the effects of this change in the flatfield are only perceptible in images with high signal-to-noise and very low natural contrast or when calculating ratios between bands. A future version of the calibration pipeline will include a variable flatfield to correct this particular artefact as well as other possible future changes in the instrument transmission.

Beside the particular case of these dust grains, comparisons between flatfields generated at different times during the mission show that differences are small. The standard deviation of the difference between the two newest products is of 0.1% and we will consider this value as the remaining pixel-to-pixel error on the flatfield frame.

Compared to the flatfield measured in the laboratory prior to flight (Roloff et al., 2017), the in-flight flatfield shows similar large-scale structures, in particular the checkboard pattern but differences at the small scale (including pixel-to-pixel variability) as well as movement of the dust particles shadows. The small-scale differences observed can be due to the effect of the radiations in space or the difference of detector temperature between the laboratory (293K) and space (273K).

*3.3. DN to I/F conversion*

The radiometric part of the data calibration – in which raw detector values (DN) are converted into physical units of radiance factor (I/F where I is the radiance collected by the instrument and F the incident irradiance from the Sun) – initially relied on a radiometric model of the instrument that is based on the assumed properties of its components, as communicated by the manufacturer (see Roloff et al., 2017 for details and Fig. 1B). In flight, images of reference stars, Phobos and Jupiter were acquired in order to verify the radiometric calibration. The comparison of these in-flight data with the model used to derive the first sets of conversion factors revealed that the actual sensitivity of the instrument is ~10% lower than expected for the PAN, RED and NIR filters, and ~5% lower than expected for BLU. As a result, the I/F values in original version of the dataset were estimated by these amounts. The entire dataset was reprocessed with the revised values and distributed during the first half of 2022. Thomas et al. (2022) detail this analysis and provide a set of multiplicative correction factors to solve this issue. The relative error on the absolute radiometry is currently estimated at 2.8%. Additional observations of reference stars will be acquired in the future to further refine these values and the full dataset will eventually be reprocessed with the final values.

Table 4 provides the values of the coefficients used to convert the instrument DN into radiance factor based on the in-flight analysis of Thomas et al. (2022). The Solar flux used in the computation of these coefficients is the one published by Meftah et al. (2018) and interpolated to a constant spectral sampling of 1 nm at a heliocentric distance of 1 Astronomical Unit. These coefficients are valid for the entire dataset. The DN values following bias subtraction and flatfield division are then multiplied by the product of these coefficients and the squared Mars heliocentric distance (in AU) and divided by the exposure time to obtain the radiance factor I/F (see equation in Fig. 3). This time- and image-dependent "absolute calibration factor" is stored in the xml header associated with each framelet. The intermediate radiance values I are not stored but can be easily recalculated from I/F if desired by using other coefficients provided in Table 4 and the squared heliocentric distance at the time of the observation. Note that we extrapolate the Sun spectrum to Mars heliocentric distance but do not take into account any Martian atmospheric influence. The I/F values derived are therefore "top-of-atmosphere" values, affected by both surface reflection and atmospheric scattering.

## 3.4. Defective pixels.

A number of pixels of the CaSSIS detector return erroneous values, either systematically or sporadically. These pixels must be identified so that the values of those which fail most of the time can be replaced by interpolation from adjacent pixels. Others should be listed in case there is doubt on the value of a single pixel while analysing images in full detail.

The detection of spurious pixels is based on a statistical analysis of the histograms of all images in the dataset. For each individual uncalibrated (level0) framelet, a histogram of raw values is produced using the IDL histogram function with a bin size of 200 DN and default values for all other parameters. Outliers to this histogram are then located. To avoid picking up pixels from the possibly discontinuous tails of the histogram, a margin corresponding to the standard deviation of the distribution of values is used. The coordinates of all pixels showing values that deviate from the boundaries of the histogram by more than one standard deviation, either below or above, are noted. For each of these identified outliers, a "percentage of failure" is then calculated for an entire boot folder, which corresponds to about one day of operation and typically several hundred framelets. This percentage is obtained by dividing the number of times a given pixel appears as an outlier and the total number of framelets in which the pixel appears. Performing this analysis with such a high temporal resolution allows us to also effectively monitor the evolution of the failure rates of bad pixels over the course of the mission.

Indeed, the failure rates of some of these defective pixels seem to evolve with time throughout the mission, as illustrated in Figure 8 with pixel [1606, 1576] (BLU filter). The evolution of the failure rate is non-systematic for each defective pixel and does not seem to correlate with any parameter of the instrument or the observations. While this has not been considered for the moment because of the low number of defective pixels found, a time-dependent update to the correction may be implemented based on the temporal analysis of the failure rates, interpolating the values of the defective pixels only for boot folders where the failure rate is higher than a chosen threshold. Importantly, no new defective pixels were identified during the first ~600 days of the mission, i.e. all known defective pixels display erratic behaviour since the beginning of the mission, and all pixels that were good at the beginning of the mission have remained so, and have not developed erratic behaviour.

While the analysis of an individual framelet with the method described here is not without potential issues and biases, the robustness of the analysis ultimately comes from the very large number of framelets on which it is applied. The failure rates provided here should be considered lower boundaries as some framelets are discarded from the analysis because of saturation and other possible issues. Finally, for all identified defective pixels, an overall failure rate is calculated by considering all failures and the total number of framelets within the entire dataset, and results are output. A table summarizing these results, including the coordinates and failure rates of all the bad pixels identified to date is provided in SOM 1. This analysis identifies 4 pixels with failure rates higher than 80%, 3 pixels with failure rates between 10 and 50%, 5 pixels with failure rates between 5 and 10%, 22 pixels with failure rates between 1 and 5% and 115 pixels that appear in the list with failure rates below 1%, most of which show less than 10 reported failures in the entire dataset. Note that pixels with a total number of failures lower than 5 were considered as false positives and discarded from the analysis.

Earlier in the mission and before the detailed analysis presented here could be performed, the decision was taken to interpolate the values of a total of 10 pixels (9 in PAN, 1 in BLU) based on the results of a preliminary version of the method used here. These interpolated pixels include the three defective pixels in PAN with failure rates higher than 90% but the bad pixel in BLU with a failure rate of ~80% was overlooked. As a result, the first version of the dataset calibrated and archived displays a bright pixel in BLU which can be very prominent in shadowed areas of images. This artefact will be removed in subsequent versions of the calibrated dataset.

While looking for potential causes for the slightly reduced sensitivity of the instrument compared to predictions (by 5 to 10%, see Thomas et al., 2022), we investigated the possibility that permanently saturated pixels affect the quantum efficiency of the detector. As we know that such dead pixels are not present in the areas of the detector used for default surface imaging, we acquired on 31-08-2020 a special dataset at a reduced cadence but with the entire detector read to make sure that no dead pixel was present on unused areas of the detector. None were found, discarding this explanation for the reduced sensitivity of CaSSIS. Additional analyses using different methods are currently performed to identify spurious behaviours by isolated pixels in the BLU filter which are missed by the current technique and possibly to understand the cause of the problem. An update to the list is therefore expected in the future.

*3.5. Straylight pattern*

From the first images returned by CaSSIS during the primary science phase, it became evident that the instrument is affected by a small and highly variable amount of straylight, with an average relative intensity of 0.3% in the PAN filter, 0.45% in BLU and much less in RED and NIR. Maxima of about 100 DN can represent up to 2% of the signal in the PAN filter depending on the illumination conditions. Due to the push-frame mechanism used by CaSSIS, and the need to mosaic consecutive framelets to generate images, even a small amount of straylight will produce obvious artefacts on the mosaicked image in the form of a repetitive banding pattern cross-track. This pattern is particularly obvious on homogeneous images with low signal, and while it might be well hidden in high-signal images of strongly contrasted scenes, advanced analyses of the latter images (e.g. colour ratios, spectral slopes, principal component analysis, etc.) often reveal the straylight pattern. Because of the importance of detailed colour analyses, including images acquired under low-light conditions, the removal of straylight, despite its low amount, is the most challenging and crucial part of the radiometric calibration of CaSSIS data.

The straylight produces a peculiar pattern on the detector with gradients at the bottom of the PAN and BLU filters and bands within the RED and NIR filters. Stacking and averaging all framelets from a given observation (as is done for producing and analysing the flatfields (Section 3.2, Figure 7)), can highlight this pattern for observations of homogeneous scenes. Comparisons between patterns from individual observations and an overall average from the entire dataset (Figure 9) reveal the high stability of the straylight pattern in time. This is key for the development of a successful correction method. With this considered, the averaged straylight pattern shown in Figure 9 was derived in the same way as the flatfield in Section 3.2 but with observations selected based on a high standard deviation along the y-axis, rather than a low standard deviation for the standard flatfield (Figure 6). The straylight pattern is the difference between this high-straylight flatfield and the standard flatfield produced previously. It must also be noted that although the spatial pattern is constant, the intensity of the straylight varies greatly with time.

Taking into consideration these observed properties of the straylight, the basis for the correction method can be outlined:

1. A constant spatial pattern is first produced from the statistical analysis of the entire dataset (see Fig. 9 and related text). This is one of the main calibration products and is archived together with the bias and flatfield frames. As for the bias and the flatfield, new versions can be produced at different times of the mission to include new observations. Note that while the first straylight pattern product had to be generated using the method described previously, posterior versions can be generated using the results of past straylight corrections, selecting observations for the high-straylight and low-straylight products based on the amount of straylight found by the correction algorithm. However, both methods were compared and little difference was found in practice.

2. A recursive algorithm determines the optimum amount of straylight to subtract from a given observation to remove the pattern (see Fig. 10 for illustration and examples). The algorithm works on an average y-profile produced by first averaging all framelets of the observation and then all pixels along the x-axis of the detector for that same given observation. The straylight pattern (fig. 9) is averaged along the x-axis in the same way. It is then stretched in steps to find the straylight intensity to remove that results in the most linear y-averaged profile. Explanations and consequences for this choice of optimization are further discussed in Section 3.6.1. Note also that because of the way two averaged products are used to generate the flatfield and straylight pattern, (using tens to hundreds of averaged observations with variable levels of straylight contained within ranges) there are many observations for which the level of straylight is actually lower than in the flatfield. In this case, the straylight correction results in an apparently negative amount of straylight. This has of course no physical meaning and only reflects the fact that the standard flatfield is contaminated by some level of straylight.

For each observation, the straylight pattern correction algorithm outputs an amplitude of the straylight pattern, in DN, defined as the maximum difference between the uncorrected and corrected averaged y-profiles. Note that this value actually constitutes a lower estimate for the actual absolute straylight, as a putative constant amount of straylight over the entire spectral window acquired would be ignored and left untouched by the correction. The only parameter that the correction allows us to refine is the relative amplitude of the straylight in the different spectral windows, as depicted in Fig. 9B. In this profile, produced for visualization only, the four segments of the curve have been arbitrarily shifted vertically (but not stretched) to visually "align" the absolute straylight levels across the detector, as a best-guess estimate of what the absolute straylight level would look like. Additional unknown offsets could apply to all four spectral windows. Because of the uncertainties in the absolute level of straylight, our straylight pattern correction is only applied in a relative way, which does not affect the absolute signal amplitude (i.e. it is applied so that the average signal in DN over the spectral window remains constant). This is currently the main limitation of this procedure and results in uncertainties on the absolute calibration of reflectance values of the order of 0.03% (RED) to 0.45% (BLU) in average over the entire dataset considering the corrected straylight pattern only. Even considering the possibility that a putative additional homogeneous background might be as high as the observed pattern, therefore doubling the error, the uncertainty would remain below 1% in the BLU filter (0.9%) and lower in other filters (0.6% in PAN, 0.35% in NIR, <0.1% in RED). Further tests in specific observations geometries will be necessary to better constrain the absolute intensity of the straylight, hence the error made of the absolute calibration and eventually fully correct it.

While the straylight pattern correction leaves small uncertainties in the absolute correction of reflectance data, it is able to efficiently remove the banding patterns from colour images as illustrated in Fig. 10 and Fig. 11 (the latter also underwent a second correction step presented in Section 3.6.1). These two examples show images of high-latitude regions taken in low-illumination conditions because of scientific constraints on the season and time of the day. This is the configuration where the effects of the correction are the most obvious as the signal is low and the level of straylight is high. In general, the correction method works until the signal level is too low to permit accurate scientific analysis of the data. On well-illuminated high contrast images, the effect of the straylight correction might not be noticeable at first glance, but more advanced analyses of colour data, such as principal component analysis, show the necessity and efficiency of the correction.

Although they are affected by uncertainties, the statistical analysis of the output straylight intensity values over the large dataset allow us to better constrain the origin of the straylight artefact. First, the level of straylight does not seem to depend on the level of signal in the image (Fig. 13A), and artefacts caused by straylight are therefore more visible on low-signal images taken at high incidence angles and/or over dark terrain. The absence of a positive correlation between the level of signal and the level of straylight indicates that the straylight is external (out-of-field). A correlation analysis of the entire dataset showed that the amount of straylight is highly dependent on both the phase angle (Fig. 13B) and the azimuth of the Sun (Fig. 13C) with respect to the telescope. This dependence on solar azimuth, (and thus on telescope orientation) is particularly obvious on many stereo pairs: one of the two stereo images tends to show a large amount of straylight, while the other image shows much less (see example in SOM 2).

All these observations point towards contributions of light reflected by the illuminated Martian surface, and possibly also the limb of the atmosphere on elements of the telescope. This suggests an insufficient baffling of the instrument.

A first potential cause of the straylight pattern is the reflection of light scattered by the illuminated disc of Mars on the back side of the sunshield, which protects the telescope from direct sunlight. This element was added late in the design, and because of mass constraints, an asymmetric design was chosen (see Thomas et al., 2017 for drawing, pictures and explanation). No straylight simulation has been performed to prove or disprove this hypothesis, however.

This task would in any case be complicated by the fact that the sunshield is entirely covered by an uneven layer of multi-layer-insulation (MLI).

Another potential contribution to the straylight pattern is suggested by CaSSIS observations of Phobos from the capture orbit of TGO (see Thomas et al., 2022). These sequences show an optical ghost (a second image of the target appearing at a different location and with a reduced intensity compared to the primary image) with an intensity of ~1% of the original signal. The ghost image of Phobos appears in the NIR filter as Phobos is leaving the field-of-view from the side of the PAN filter (Fig. 14). The ghost image is almost in focus and the direct and ghost images move in opposite directions. No ghost is observed on the other side of the detector (BLU filter). Using these observations, we suspect that the ghost image originates from a specular reflection on the side-face of a carbon fibre piece used as a field stop at the intermediate focus of the telescope. This optical ghost cannot explain the entirety of the out-of-field straylight observed by CaSSIS as we would expect a positive correlation between signal and straylight in this case but it might contribute to it.

The insufficient baffling of the aperture and internal reflections will be corrected for in subsequent iterations of this telescope used in future instruments. For instance, a cylindrical baffle instead of the asymmetric sunshield, and a knife-edge profile for the field stop are now being considered for the telescope of the Comet Camera (CoCa) imager developed for the Comet Interceptor mission, which is based on an updated CaSSIS design.

### *3.6. Gradients and framelet-to-framelet offsets*

3.6.1. Residual gradients from the straylight pattern correction

We explained in Section 3.5 that the straylight pattern correction method involves an optimization step to remove the straylight pattern from the y-averaged profile so that the resulting y-profile is as linear as possible (possibly with a slope), and not as flat (i.e. horizontal) as possible (See Fig. 9). This is because actual gradients along the y-axis of the detector can exist in images that show a brightness gradient along the along-track dimension of the image. Trying to correct for such gradients results in the introduction of visible artefacts into the images instead of a removal. In other cases, the pattern removal leaves behind linear gradients which are clear artefacts, since they are seen to differ between colours. It is likely that in addition to the constant pattern, the straylight also has a variable component in the form of a linear gradient across the detector which produces the effects observed.

These gradients can be easily analysed and retrieved from the data by computing the differences of signal in overlapping areas between adjacent framelets. When plotted for all framelets and all colours from a given observation, the signal difference between adjacent framelets will show constant values for all framelets and for each colour but might be different between colours. Compared to the intensity of the pattern removed in the previous step, these gradients are smaller, typically of the order of a few DN.

3.6.2. Framelet-to-framelet offsets

Besides the straylight pattern discussed in Section 3.5, sudden brightness offsets between framelets are a rarer but quite prominent artefact seen on some CaSSIS images. Examples are in Fig. 15 and Fig 16, which show, respectively one section of an image that has a distinctive yellow colour compared to the rest of the scene, and a section of another image featuring icy deposits that have a purple hue, while the ice looks white in the rest of the image. As for the straylight pattern, these artefacts are more noticeable on images with low signal or low contrast and often appear in images of homogeneous scenes of dusty and smooth plains (Fig. 15), or images taken at high latitude (Fig. 16).

This artefact is analysed in the same way as the gradients described in Section 3.6.1, i.e. by computing the difference of signal between overlapping areas of adjacent framelets. The analysis shows that the sudden offsets occur simultaneously in all colours. Note that because the images acquired at the same time through different filters on the detector correspond to different areas of the surface (Fig. 1), the artefact will appear at different locations on the different colour images. In addition to appearing simultaneously on all filters, the artefact also shows the same absolute amplitude, independent of the difference in absolute signal between the data of different filters. This clearly indicates that the offsets are caused by variations in the bias level of the detector. This was also confirmed by the analysis of dark frames taken simultaneously with the image frames in a few cases. As shown in Fig. 1, two small control windows (Ctrl. win #1 and Ctrl. win #2) were defined on opposite corners of the detector below the black mask that covers unused areas of the detector. These can be read simultaneously with the actual imaging data, and their purpose was indeed to check for the stability of the bias during a series of acquisitions. The control windows were included because testing of the flight detector in the lab (Roloff et al., 2017), as well as similar investigations on the twin SIMBIO-SYS detectors (Simioni et al., 2016; Slemer et al., 2019) pointed toward potential issues related to bias stability. In flight, these control windows have only been used sporadically, as data transmission bottlenecks in the instrument mean that their use slightly restricts the size of the images acquired. In cases in which they were used, it is possible to compare the outputs from the control windows with the results of the method described previously. Fig. 17 demonstrates four different examples in which both methods give very similar results, justifying the use of the method based on the overlaps but also demonstrating the potential usefulness of the control windows in cases where they could be more suitable. This exercise also proves that the observed artefacts are indeed caused by sudden changes in the detector bias. Another approach was successfully tested during the ICO4 commissioning phase of the STC instrument of SIMBIO-SYS, which shares the same detector with CaSSIS. The test demonstrates that a limited number of acquisitions of a small 64x128 pixels window before the science acquisitions strongly mitigates this effect. The additive DC level was reduced by a factor of 10 with 2 acquisitions of the small windows and by 100 with 4 acquisition of small windows.

The values of the corrected framelet-to-framelet bias offsets are output by the calibration pipeline and stored in report files. These values can then be analysed statistically to look for possible causes of the bias artefact. The only correlation found was with the actual signal of the image (Fig. 18). The higher the signal in the image, the larger the offsets can be. Images with low signal usually show no or very limited offsets. The detector nearing saturation (but not reaching it) therefore seems to be the dominant cause for these artefacts.

3.6.3. Strategy and correction algorithm

Both artefacts discussed in this section are characterized using the same technique: it relies on comparisons of the signal in overlapping areas between adjacent framelets. They are also corrected together, at the same step of the calibration pipeline and as illustrated in Fig. 19. Once the signal differences have been calculated for all pairs of framelets and all colours, the gradients are removed. The median of all differences is calculated independently for each filter, which captures the residual gradients described in Section 3.6.1 while limiting the influence of the outliers from the sudden offsets described in Section 3.6.2. A linear gradient across the y-dimension of the detector with the amplitude of this median difference value is then subtracted from all data (Fig. 19, A to B). The main assumption made here is that the gradients are purely linear and any non-linearity would remain uncorrected.

In a second step, the residual offsets are averaged for all colours acquired, and subtracted (Fig. 19, B to C). Averaging for all colours improves the accuracy, especially for the BLU, RED and NIR filters for which the signal is lower. This method efficiently corrects the sudden bias changes described in Section 3.6.2 as illustrated in Figs. 15 and 16. Note that because of the averaging between colours, this correction step works better for images where all four colour filters are acquired.

## 4. Summary and Discussion

The CaSSIS instrument was developed on a tight schedule with the objective of serving as the scientific imager on board ExoMars TGO. The condensed schedule meant that the testing phase prior to launch was very short and only a reduced number of tests and calibrations were feasible. However, these tests were sufficient to verify the basic functionality of the instrument and qualify it for launch and operations in space (Roloff et al., 2017). Naturally, it was already clear that additional in-flight characterizations and calibrations would be needed for a proper understanding and use of the instrument and its data.

In flight, the instrument demonstrated outstanding performance and provided sharp images, high SNR, and excellent colour sensitivity, as demonstrated by more than 20,000 acquired images to date, most of them of very high quality. The first years of operations have also been complicated by software bugs and technical glitches, which were all corrected with a flight software update in July 2020. The operations of the instrument are now smoother than they have ever been in the mission, with a highly-optimized use of the data volume and rare loss of images (see Almeida et al., this issue). Although the use of the rotation mechanism is restricted compared to the beginning of the mission, it is still functional and allows the planning of regular stereo images, and aligning of the instrument to the ground-track for perfect colour overlap, albeit with lower frequency than originally planned.

The radiometric calibration of images improved regularly during the first years of the primary science mission, as better calibration products were produced from in-flight data and replaced the products initially derived from pre-flight laboratory data. The push-frame approach used by CaSSIS proved to be extremely beneficial in this regard, as the high number of images taken at high cadence allowed averaging and statistical study of the signal, to produce improved bias and flatfield frames from carefully filtered and selected data.

Both the detector and the telescope suffer from technical issues which result in visible artefacts on uncorrected data. The bias level of the detector becomes unstable when high levels of signal are recorded in large areas of the detector, even before reaching saturation. This produces additive offsets between adjacent framelets (of the same amplitude for all colours). Internal reflections within the telescope and/or its sunshield produce out-of-field straylight with a fixed spatial pattern, but highly variable amplitude depending on the position of the Sun and illuminated Mars surface. This results in stripes and gradients across framelets which produce repetitive patterns across the entire image.

The amplitudes of both types of artefacts are relatively small; of the order of a few percent of the total signal at maximum. However, because of the way colour images are generated (mosaicking tens of framelets), the effects of the artefacts are apparent on images where the signal is low, and even if they are not on high-signal images, they become obvious when the colours are analysed with advanced techniques. Fortunately, the push-frame approach, which makes the artefacts prominent in the end-products, also results in advantages that help correct them. Indeed, for each image that CaSSIS takes, the 30 to 50 framelets that are successively acquired can be stacked, averaged and their standard deviation calculated. This allows for averaging of the surface signal to retrieve the amplitude of the artefacts in a precise manner. This is a fundamental step of the correction method for the straylight pattern, in which the amplitude of a fixed straylight pattern is adjusted to each individual image. The overlap between adjacent framelets permits the retrieval of very small changes in the bias of the detector (at a level of only a few DN) and the correction of residual gradients left by the previous step of the straylight correction.

In essence, while the push-frame approach is more sensitive to these two types of artefacts than other approaches, it also allows us to correct them more precisely. Indeed, a few percent of straylight across the detector, and changes in detector bias of up to a few tens of DN could certainly go undetected in a framing camera and induce error in the physical calibration.

While our radiometric corrections were shown to be efficient in most cases, the non Sun-synchronous orbit of TGO permits imaging of the surface under very variable illumination conditions, from close to opposition to directly over the terminator. Because of this configuration, some planning periods only allow the selection of targets at very high phase angle. As a result, the CaSSIS dataset has a very wide range of illumination conditions, and hence a highly variable signal and SNR. Figure 21 presents an analysis of the SNR in all four filters as a function of the solar incidence angle. These plots clearly show two regimes: at lower incidence angles, the exposure time can be adjusted to values lower than the 1-pixel smearing time to target a mean signal of ~8000 DN in PAN, which corresponds to about 2/3 of the detector's full-well depth. This is the optimal configuration that maximizes the signal, minimizing the photon noise while mitigating risks of local saturation over bright areas of the image. Knowing that the signal is about two times lower in RED and NIR and four times lower in BLU, this translates to values of SNR of ~230 for PAN, ~160 for RED and NIR and ~110 for BLU in this regime (the photon noise is the square root of the signal, and a readout noise of 8 DN is considered). Images acquired in these conditions represent about 60% of the entire dataset.

While these values might seem high, it is important to consider that the dusty atmosphere of Mars has a strong effect on the surface signal. Under normal atmospheric conditions, about half of the photons scattered back into space by the surface also interact with aerosols, strongly reducing the contrast of surface features. Because atmospheric conditions vary drastically spatially and temporally, the SNRs of CaSSIS images also show large variations in quality. In extreme PEDE such as the one of July 2018, imaging is fully halted, as the surface is no longer visible to the instrument.

At incidence angles larger than ~60°, the SNR decreases as incidence angle increases. As the incoming solar flux decreases proportionally to the cosine of the incidence angle, the optimal exposure time becomes larger than the 1-pixel smearing time, which therefore limits the exposure duration. There is a large range of incidence angles at which the transition between the two regimes occurs due to the variability in surface albedo and the change of the heliocentric distance with time, since the orbit of Mars is eccentric. With a clear atmosphere and over bright surfaces, CaSSIS can still acquire high-quality colour images of the surface at incidence angles up to about 75 – 80° with an SNR of ~150 in PAN, ~100 in RED and NIR and ~70 in BLU. This is the case of the images shown in Fig. 11 and Fig. 12, where the incidence angle reaches 76°.

For incidence angles larger than ~80° (cos i = 0.17), the signal from the surface decreases drastically and the contribution of atmospheric aerosols dominates the signal. Surface features can still be recognized and analysed in PAN, but most of the colour information is lost in the noise. At this point, the radiometric correction also shows its limits, and residuals from the straylight and offset corrections (Table 5) can remain prominent. Fig. 21 A) and B) show the example of an image acquired with an incidence angle of 87° (the Sun is only 3° above the horizon) which illustrates this case. Because of the high incidence angle, the measured median signal in the three filters used in this image were 1200 DN in PAN, 500 DN in NIR, and 400 DN in BLU, dominated by atmospheric scattering in the latter case. Surface features are recognizable and can be analysed thanks to the higher SNR in the PAN filter. There is however no colour information, and zooming in on the surface only reveals noise in the NIR and BLU filters. An insufficient correction of the straylight results in regular brightness and colour patterns on both sides of the image, but the central part is better corrected. It is also noticeable here that the left portion of the image shows a distinctive reddish tone, while the right portion instead appears bluish. The cause of this latter artefact is not yet identified but it must be related to the detector reading electronics as it affects exactly half of the detector. In addition to low-signal images acquired at high incidence angle, this artefact is also visible sometimes albeit quite subtle, in high-signal images of very homogenous scenes (see an example in SOM 3). We plan to further analyse this particular artefact to try to identify its cause and possibly correct at least part of it.

Because of the strong contrast of reflectance at blue wavelengths between ice and ice-free surfaces, images taken at very high incidence angle can still be very valuable in helping to map frost and ice deposits in low-illumination conditions, such as early local times and early spring, or late autumn at high latitudes. Fig. 21 C) and D) illustrate this case through the example of the rim of an unnamed crater at high northern latitude. The image was taken on the morning terminator with an incidence angle of exactly 90°. Despite the very low signal (500 DN in PAN, 240 in RED and 210 in BLU) – and hence the very low SNR (42 in PAN, 23 in RED, 21 in BLU) – the patches of ice on the rim can be clearly identified through the contrast of reflectance in PAN, where the SNR is highest. Further, they can also be identified through the difference of spectral slope between the red dust and the white ice, which is discernible even in these difficult conditions.

The THEMIS VIS instrument of Mars Odyssey was the first imager to implement a push-frame approach for visible colour imaging of the Martian surface. Interestingly, similar challenges were encountered during the calibration of this instrument (McConnochie et al., 2006). Significant out-of-field straylight was also a major concern, but it could ultimately be corrected to retrieve reflectance values with a good 2σ accuracy, estimated to be better than 5% at 654 nm. We note however that the type of straylight that affects THEMIS VIS is fundamentally different from the one affecting CaSSIS, as the straylight signal is proportional to the scene radiance for THEMIS, whereas it is independent of it for CaSSIS.

Mahanti et al. (2016) detail the in-flight calibration of the Lunar Reconnaissance Orbiter Wide Angle Camera (LRO-WAC), another push-frame visible colour camera. For this instrument, the bias and dark levels are the major concern, in terms of calibration, since they depend on temperature and exposure time. Dark observations must therefore be acquired regularly and interpolated. However, LRO-WAC is equipped with a very large baffle to protect it against out-of-field straylight, and thus the straylight performance of the instrument is good and no correction had to be implemented. As for CaSSIS, an optical ghost was found with an intensity of about 1% of the maximum signal strength.

Because CaSSIS shares some of the same heritage as the SIMBIO-SYS colour imager on BepiColombo, many of these lessons learned with the operation and calibration of CaSSIS will be of immediate interest to SIMBIO-SYS team. Indeed, the instruments utilise a push-frame approach and share an identical detector with colour filters applied directly above the detector. In fact, these lessons are already being applied. For instance, the variable bias of the detector was identified and studied during the ground calibration of SIMBIO-SYS and linked to an inefficiency in the RESET function of the sensor. As with CaSSIS (though only occasionally used (Figure 17)), control windows behind the black mask around the colour filters will be defined and read to characterize the changes of dark signal. Nevertheless, the optical design of the SIMBIO-SYS telescope is very different and the other artefacts encountered with CaSSIS may not be relevant with the SIMBIO-SYS imager.

# 5. Conclusion

With more than 30,000 images of the Martian surface acquired during its first four years of operations, CaSSIS has collected a valuable dataset which complements results from other instruments and missions to decipher the geological history of the red planet. The CaSSIS calibration products and procedure were updated in-flight, using both standard and specific observations conducted from the nominal science mission orbit. This was especially necessary due to a tight development schedule that limited laboratory tests and calibrations, however, the space environment always affects instrument performance, and these improvements would always have been performed and proved useful. CaSSIS uses a push-frame approach for colour imaging which results in both advantages and challenges in terms of radiometric calibration. As every image is actually a mosaic of several tens of small images called framelets, any issue with the relative radiometric calibration of the individual framelets immediately appears as a repetitive pattern on the final mosaic. However, this means that for each image, all framelets can be averaged and analysed statistically to retrieve the amplitude of the radiometric artefacts independently of the surface signal. Overlapping areas between adjacent framelets can also be used to track erratic changes in the signal levels. Such corrections were necessary as CaSSIS suffers from two distinct technical issues that combine to affect its images: (1) Insufficient baffling and internal reflections in the telescope result in out-of-field straylight with up to a couple of % of the actual signal intensity, and (2) Erratic variations in the bias level of the sensor also result in offsets between adjacent framelets of the same amplitude. Both issues are corrected efficiently using the procedures detailed in this paper. In the same way, the multitude of acquisitions necessary to produce the images is beneficial for the production of dark and flatfield frames with good SNR. We have compared here the calibration products and procedures currently in use in the calibration pipeline with more recent versions. Overall, we can conclude that changes are small and improvements will not be further significant (e.g. a few bad pixels could have been overlooked, or one dust particle could have moved in the telescope, etc.). We will nevertheless continue to monitor changes and update products so that future calibrations of the dataset are up-to-date with our knowledge of the instrument.

## Acknowledgements

CaSSIS is a project of the University of Bern and funded through the Swiss Space Office via ESA's PRODEX programme. The instrument hardware development was also supported by the Italian Space Agency (ASI) (ASI-INAF agreement no. 2020-17-HH.0), INAF/Astronomical Observatory of Padova, and the Space Research Center (CBK) in Warsaw. Support from SGF (Budapest), the University of Arizona (Lunar and Planetary Lab.) and NASA are also gratefully acknowledged. Operations support from the UK Space Agency under grant ST/R003025/1 is also acknowledged. LLT wishes to personally acknowledge funding and support from the Canadian Space Agency (CSA) through their Planetary and Astronomy Missions Co-Investigator programme (22EXPCOI3) and the Canadian NSERC Discovery Grant programme (RGPIN 2020-06418).  We are grateful to A. Hayes and an anonymous reviewer for their attentive reading of the manuscript and their constructive remarks and comments which were very useful to improve the manuscript.

# References

Almeida, M., Read, M., Thomas, N., Cremonese, G., Becerra, P., Borrini, G., Gruber, M., Heyd, R., Marriner, C.M., McArthur, G., McEwen, A.S., Pommerol, A., Perry, J., Schaller, C, Targeted Planning on Mars with ExoMars/CaSSIS, submitted to PSS (this issue)

Anderson, J. A., & Robinson, M. S., Challenges Utilizing Pushframe Camera Images, Lunar and Planetary Science Conference, 1905, (2009).

Besse, S., Vallat, C., Barthelemy, M., Coia, D., Costa, M., De Marchi, G., Fraga, D., Grotheer, E., Heather, D., Lim, T., Martinez, S., Arviset, C., Barbarisi, I., Docasal, R., Macfarlane, A., Rios, C., Saiz, J., & Vallejo, F., ESA's Planetary Science Archive: Preserve and present reliable scientific data sets, Planetary and Space Science, 150, 131, (2018).

Christensen, P. R., Jakosky, B. M., Kieffer, H. H., Malin, M. C., McSween, H. Y., Nealson, K., Mehall, G. L., Silverman, S. H., Ferry, S., Caplinger, M., & Ravine, M., The Thermal Emission Imaging System (THEMIS) for the Mars 2001 Odyssey Mission, Space Science Reviews, 110, 85, (2004).

Cremonese, G., Capaccioni, F., Capria, M. T., Doressoundiram, A., Palumbo, P., Vincendon, M., Massironi, M., Debei, S., Zusi, M., Altieri, F., Amoroso, M., Aroldi, G., Baroni, M., Barucci, A., Bellucci, G., Benkhoff, J., Besse, S., Bettanini, C., Blecka, M., Borrelli, D., Brucato, J. R., Carli, C., Carlier, V., Cerroni, P., Cicchetti, A., Colangeli, L., Dami, M., Da Deppo, V., Della Corte, V., De Sanctis, M. C., Erard, S., Esposito, F., Fantinel, D., Ferranti, L., Ferri, F., Ficai Veltroni, I., Filacchione, G., Flamini, E., Forlani, G., Fornasier, S., Forni, O., Fulchignoni, M., Galluzzi, V., Gwinner, K., Ip, W., Jorda, L., Langevin, Y., Lara, L., Leblanc, F., Leyrat, C., Li, Y., Marchi, S., Marinangeli, L., Marzari, F., Mazzotta Epifani, E., Mendillo, M., Mennella, V., Mugnuolo, R., Muinonen, K., Naletto, G., Noschese, R., Palomba, E., Paolinetti, R., Perna, D., Piccioni, G., Politi, R., Poulet, F., Ragazzoni, R., Re, C., Rossi, M., Rotundi, A., Salemi, G., Sgavetti, M., Simioni, E., Thomas, N., Tommasi, L., Turella, A., Van Hoolst, T., Wilson, L., Zambon, F., Aboudan, A., Barraud, O., Bott, N., Borin, P., Colombatti, G., El Yazidi, M., Ferrari, S., Flahaut, J., Giacomini, L., Guzzetta, L., Lucchetti, A., Martellato, E., Pajola, M., Slemer, A., Tognon, G., & Turrini, D., SIMBIO-SYS: Scientific Cameras and Spectrometer for the BepiColombo Mission, Space Science Reviews, 216, 75, (2020).

Gambicorti, L., Piazza, D., Gerber, M., Pommerol, A., Roloff, V., Ziethe, R., Zimmermann, C., Da Deppo, V., Cremonese, G., Ficai Veltroni, I., Marinai, M., Di Carmine, E., Bauer, T., Moebius, P., & Thomas, N., Thin-film optical pass band filters based on new photo-lithographic process for CaSSIS FPA detector on Exomars TGO mission: development, integration, and test, Proceedings of the SPIE, Volume 9912, id. 99122Y 8 pp. (2016).

Gambicorti, L., Piazza, D., Pommerol, A., Roloff, V., Gerber, M., Ziethe, R., El-Maarry, M. R., Weigel, T., Johnson, M., Vernani, D., Pelo, E., Da Deppo, V., Cremonese, G., Ficai Veltroni, I., & Thomas, N., First light of Cassis: the stereo surface imaging system onboard the exomars TGO, Society of Photo-Optical Instrumentation Engineers (SPIE) Conference Series, 10562, 105620A, (2017).

Mahanti, P., Humm, D. C., Robinson, M. S., Boyd, A. K., Stelling, R., Sato, H., Denevi, B. W., Braden, S. E., Bowman-Cisneros, E., Brylow, S. M., & Tschimmel, M., Inflight Calibration of the Lunar Reconnaissance Orbiter Camera Wide Angle Camera, Space Science Reviews, 200, 393, (2016).

McConnochie, T. H., Bell, J. F., Savransky, D., Mehall, G., Caplinger, M., Christensen, P. R., Cherednik, L., Bender, K., & Dombovari, A., Calibration and in-flight performance of the Mars Odyssey Thermal Emission Imaging System visible imaging subsystem (THEMIS VIS), Journal of Geophysical Research (Planets), 111, E06018, (2006).

Perry, J. E., Heyd, R., Read, M., Tornabene, L. L., Sutton, S., Byrne, S., Thomas, N., Fennema, A., McEwen, A., and Berry, K., Geometry Pipeline for TGO CaSSIS Observations, submitted to PSS (submitted to PSS, this special issue).

Planetary Data System, Planetary Data System Standards Reference pds4. https://pds.nasa.gov/pds4/doc/sr/. Version 1.7.0, 2016.

Re, C. Fennema, A., Simioni, E., Sutton, S., Mège, D., Gwinner, K., Józefowicz, M., Munaretto, G., Petrella, A., Pommerol, A., Cremonese, G., Thomas, N., CaSSIS-based stereo products for Mars after three years in orbit, submitted to PSS (this issue).

Robinson, M. S., Brylow, S. M., Tschimmel, M., Humm, D., Lawrence, S. J., Thomas, P. C., Denevi, B. W., Bowman-Cisneros, E., Zerr, J., Ravine, M. A., Caplinger, M. A., Ghaemi, F. T., Schaffner, J. A., Malin, M. C., Mahanti, P., Bartels, A., Anderson, J., Tran, T. N., Eliason, E. M., McEwen, A. S., Turtle, E., Jolliff, B. L., & Hiesinger, H., Lunar Reconnaissance Orbiter Camera (LROC) Instrument Overview, Space Science Reviews, 150, 81, (2010).

Roloff, V., Pommerol, A., Gambicorti, L., Servonet, A., Thomas, N., Brändli, M., Casciello, A., Cremonese, G., Da Deppo, V., Erismann, M., Ficai Veltroni, I., Gerber, M., Gruber, M., Gubler, P., Hausner, T., Johnson, M., Lochmatter, P., Pelò, E., Sodor, B., Szalai, S., Troznai, G., Vernani, D., Weigel, T., Ziethe, R., & Zimmermann, C., On-Ground Performance and Calibration of the ExoMars Trace Gas Orbiter CaSSIS Imager, Space Science Reviews, 212, 1871, (2017).

Simioni, E., Re, C., Mudric, T., Cremonese, G., Tulyakov, S., Petrella, A., Pommerol, A., & Thomas, N., 3DPD: A photogrammetric pipeline for a PUSH frame stereo cameras, Planetary and Space Science, 198, 105165, (2021).

Slemer, A., Da Deppo, V., Simioni, E., Re, C., Dami, M., Borrelli, D., Veltroni, I. F., Aroldi, G., Tommasi, L., Capria, M. T., Naletto, G., Mugnuolo, R., Amoroso, M., & Cremonese, G., Radiometric calibration of the SIMBIO-SYS STereo imaging Channel, CEAS Space Journal, 11, 485, (2019).

Thomas, N., Cremonese, G., Ziethe, R., Gerber, M., Brändli, M., Bruno, G., Erismann, M., Gambicorti, L., Gerber, T., Ghose, K., Gruber, M., Gubler, P., Mischler, H., Jost, J., Piazza, D., Pommerol, A., Rieder, M., Roloff, V., Servonet, A., Trottmann, W., Uthaicharoenpong, T., Zimmermann, C., Vernani, D., Johnson, M., Pelò, E., Weigel, T., Viertl, J., De Roux, N., Lochmatter, P., Sutter, G., Casciello, A., Hausner, T., Ficai Veltroni, I., Da Deppo, V., Orleanski, P., Nowosielski, W., Zawistowski, T., Szalai, S., Sodor, B., Tulyakov, S., Troznai, G., Banaskiewicz, M., Bridges, J. C., Byrne, S., Debei, S., El-Maarry, M. R., Hauber, E., Hansen, C. J., Ivanov, A., Keszthelyi, L., Kirk, R., Kuzmin, R., Mangold, N., Marinangeli, L.,

Markiewicz, W. J., Massironi, M., McEwen, A. S., Okubo, C., Tornabene, L. L., Wajer, P., & Wray, J. J., The Colour and Stereo Surface Imaging System (CaSSIS) for the ExoMars Trace Gas Orbiter, Space Science Reviews, 212, 1897, (2017).

Thomas, N., Pommerol, A., Almeida, M., Read, M. Cremonese, G., Simioni, E., Munaretto, G., and Weigel, T., Absolute calibration of the Colour and Stereo Surface Imaging System (CaSSIS), PSS, 211, 105394, (2022).

Tulyakov, S., Ivanov, A., Thomas, N., Roloff, V., Pommerol, A., Cremonese, G., Weigel, T., & Fleuret, F., Geometric calibration of Colour and Stereo Surface Imaging System of ESA's Trace Gas Orbiter, Advances in Space Research, 61, 487, (2018).

Zusi, M., Simioni, E. T., Cicchetti, A. , Politi, R., Noschese, R., Carlier, V., Da Deppo, V., Filacchione, G., Re, C., Tommasi, L., Debei, S., Fonte, S., Langevin, Y., Slemer, A., Baroni, M., Borelli, D., Dami, M., Ficai Veltroni, I., Amoroso, M., Longo, F., Mugnuolo, R., Capaccioni, F., Capria, M. T., Palumbo, P., Vincedon, M., & Cremonese, G., SIMBIO-SYS Near Earth Commissioning Phase: a step forward toward Mercury, Infrared Remote Sensing and Instrumentation XXVII, 11128, (2019).

# Tables

| Orbit | Quasi-circular (365 x 420 km) |
|---|---|
| Orbital period | 1.96 hours |
| Orbital inclination | 74° |
| Ground speed | ~ 3 km/s |
| | |
| Telescope design | Modified Three Mirrors Anastigmat (TMA) |
| Focal length | 875±2 mm |
| Aperture | 135 mm |
| f/number | 6.5 |
| | |
| Detector type | Raytheon Osprey 2k Si PIN CMOS |
| Sensor size | 2048 x 2048 pixels |
| Sensor pixel pitch | 10 µm x 10 µm |
| Pixel readout rate | 5 MHz |
| Readout noise | 61 $e^-$ (9 DN) |
| Gain | 7.1 $e^-$/DN |
| Peak quantum efficiency | >99% (~750nm) |
| Dark current (273K) | 5000 $e^-$/s |
| Full-well capacity | 90'000 electrons |
| Digital signal resolution | 14 bit |
| | |
| Filters central wavelengths (nm, BLU, PAN, RED, NIR) | 495, 678, 836, 939 |
| Filters widths (FWHM, nm, BLU, PAN, RED, NIR) | 134, 232, 99, 122 |
| | |
| Typical imaging repetition | 350 ms |
| Exposure time | ~ 0.6 to 1.5 ms |
| Typical dimensions of an image | 9 x 40 km |
| Field of view (across track) | 1.336° |
| Angular scale | 11.36 µrad/pixel |
| Ground resolution | 4.5 m |
| | |
| Mass | 18.05 kg |
| Power (survival, average, peak) | 8.5, 17.3, 56.7 W |

**Table 1**: List of the most relevant characteristics of the CaSSIS instrument and the spacecraft orbit.

| Filter | Image ID | Incidence angle [°] | Phase angle [°] | Median signal [DN] |
|---|---|---|---|---|
| BLU | MY34_003508_125_0 | 143.37 | 138.85 | 3754.6 |
| BLU | MY34_003411_101_0 | 129.06 | 127.23 | 3754.7 |
| BLU | MY34_003502_111_0 | 137.18 | 130.83 | 3754.6 |
| BLU | MY34_003433_119_0 | 131.94 | 131.28 | 3757.6 |
| BLU | MY34_003382_098_0 | 127.53 | 127.33 | 3757.9 |
|  |  |  |  |  |
| PAN | MY34_003502_111_0 | 137.18 | 130.83 | 3812.9 |
| PAN | MY34_003418_104_0 | 129.86 | 127.99 | 3814.3 |
| PAN | MY34_003461_111_0 | 134.20 | 130.42 | 3817.5 |
| PAN | MY34_003518_138_0 | 147.27 | 144.74 | 3821.0 |
| PAN | MY34_003435_122_0 | 131.95 | 131.67 | 3821.4 |
|  |  |  |  |  |
| RED | MY34_003526_112_0 | 139.14 | 131.50 | 3772.4 |
| RED | MY34_003508_125_0 | 143.37 | 138.85 | 3773.3 |
| RED | MY34_003411_101_0 | 129.06 | 127.23 | 3773.3 |
| RED | MY34_003481_109_0 | 135.31 | 130.09 | 3774.1 |
| RED | MY34_003394_103_0 | 128.10 | 127.80 | 3781.6 |
|  |  |  |  |  |
| NIR | MY34_003433_119_0 | 131.94 | 131.28 | 3758.9 |
| NIR | MY34_003382_098_0 | 127.53 | 127.33 | 3759.5 |
| NIR | MY34_003526_112_0 | 139.14 | 131.50 | 3763.8 |
| NIR | MY34_003481_109_0 | 135.31 | 130.09 | 3765.1 |
| NIR | MY34_003418_104_0 | 129.86 | 127.99 | 3765.7 |

**Table 2**: List of CaSSIS observations used to generate the bias product currently in use in the radiometric pipeline. The selection of these observations is based on the 5 lowest values of the median signal [DN].

|            | BLU  | PAN  | RED | NIR |
|------------|------|------|-----|-----|
| "2018-08-10" | 38   | 232  | 63  | 26  |
| "2018-10-09" | 200  | 561  | 111 | 65  |
| "2019-03-08" | 756  | 2197 | 121 | 114 |
| "2019-12-20" | 934  | 2665 | 150 | 228 |
| "2020-07-27" | 1431 | 4761 | 268 | 284 |

**Table 3**: Number of CaSSIS observations averaged to generate different versions of the flatfield frame used in the radiometric calibration pipeline. The product currently in use is "2019-03-08".

| Filter | BLU | PAN | RED | NIR |
|---|---|---|---|---|
| Conversion coefficient (reflectance /(DN.s$^{-1}$)) x 10$^{-8}$ | 2.793 | 1.481 | 3.857 | 3.975 |
| I/F to rad | 1.69 | 2.34 | 3.32 | 4.09 |

**Table 4**: Coefficients used for the conversion of the instrument generated data number (DN) to radiance factor I/F (first line) and back to the physical unit of radiance (W.m$^{-2}$.Sr$^{-1}$.nm$^{-1}$) (second line).

| Product / calibration step | Estimated uncertainty |
|---|---|
| Bias frame | 0.2% overall value |
| Flatfield frame | 0.1% pixel-to-pixel overall, up to 0.5% locally |
| Absolute radiometry | 2.8% |
| Contribution of straylight to absolute signal | < 1% in average |
| Optical ghost | ~1% locally |
| Remaining straylight | Up to ~20 DN |
| Remaining offsets/gradients between framelets | Up to ~20 DN |

**Table 5**: Estimated uncertainties on calibration products and procedures

# Figures captions

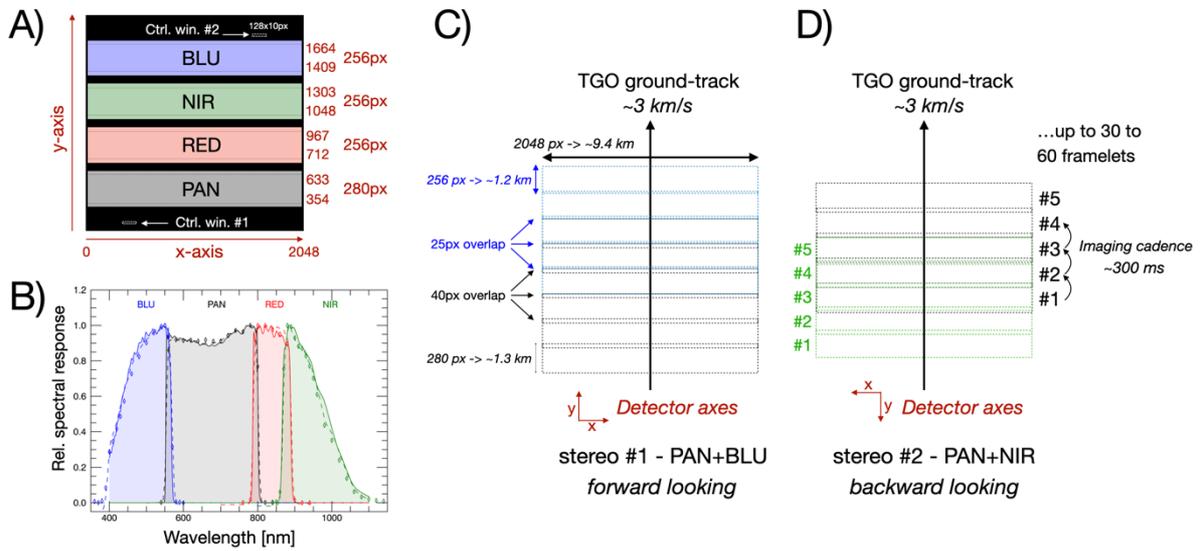

**Figure 1:** Illustration of the colour and stereo imaging principle of CaSSIS implementing a "push-frame" approach. A) Four colour bandpass filters are mounted directly on top of the detector with thin black masks in between the filters, below and above. Up to four windows which are 280px-high for the panchromatic (PAN) filter and 256px-high for the three colour filters can be read, close to the centres of the colour filters. The default vertical coordinates are indicated. In addition, two small "control windows" are defined below the black mask to read the bias/dark in unilluminated areas. B) Relative spectral response in the 4 colour filters (adapted from Roloff et al., 2017) from a model of the instrument (solid lines) compared to verification measurements in the lab prior to flight (symbols and dashed lines). C) and D) illustration of the footprints of CaSSIS individual acquisitions ("framelets") in the case of a stereo acquisition with the PAN filters on both sides of the stereo pair and the BLU and NIR filter on the forward and backward-looking side of the stereo pair, respectively.

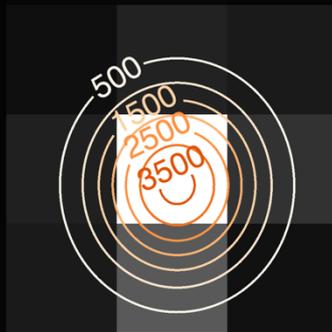

**Figure 2**: Example of the Point Spread Function (PSF) of the standard photometric star Pi$^2$ Orionis measured by CaSSIS during its first on-orbit commissioning phase in October 2016.

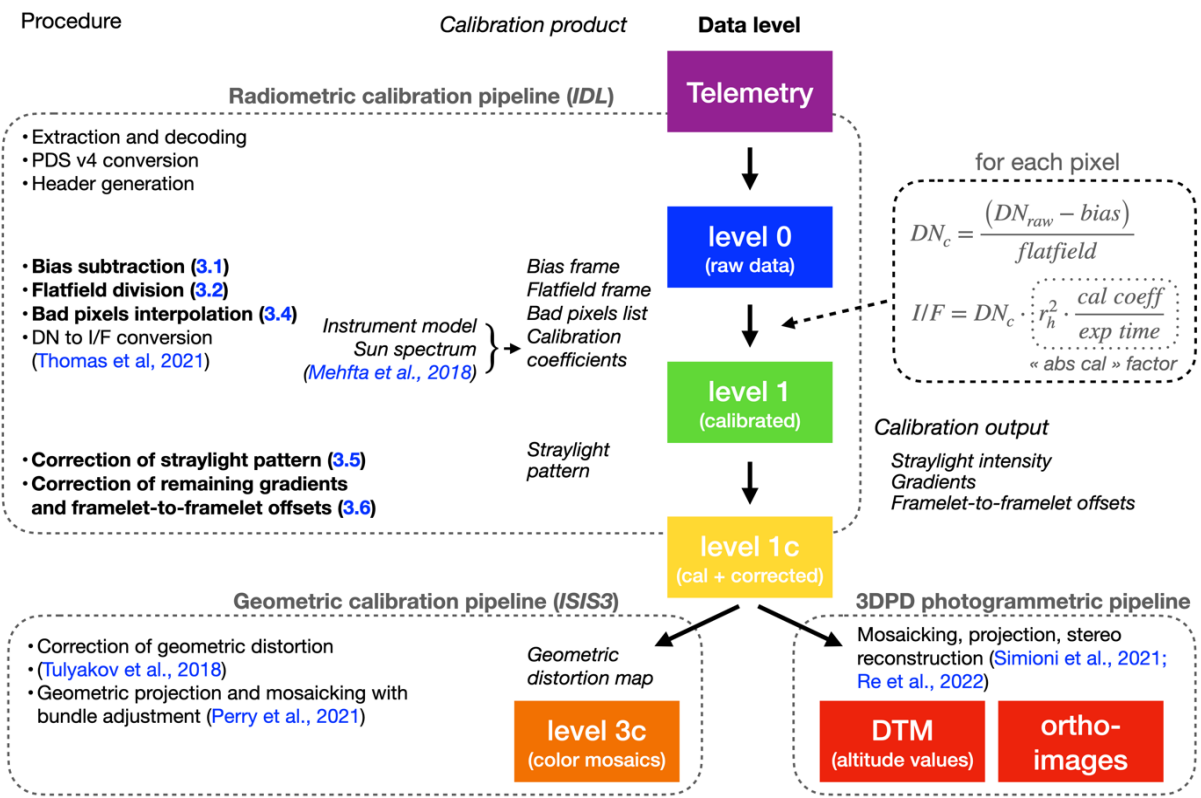

**Figure 3**: Overview of the entire calibration process, from the raw telemetry received from the spacecraft to the final archived calibrated product.

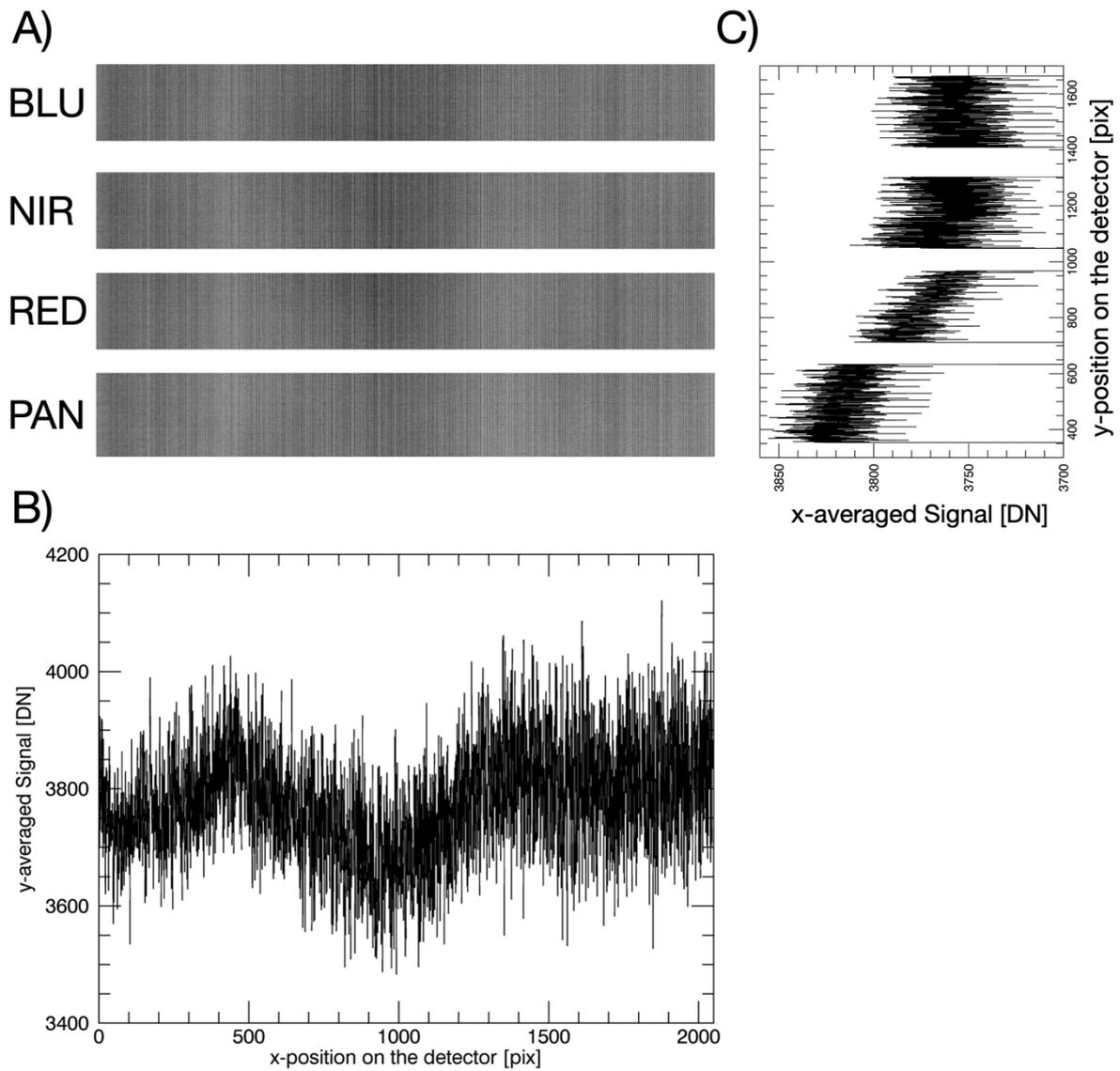

**Figure 4:** Structure of the bias frame used for the calibration of all CaSSIS images. This bias is subtracted from all raw images, which represents the first step of the photometric calibration procedure. A) Image of the bias frame linearly stretched between DN values 3200 and 4500. B) and C) average profiles along the x- and y- directions of the detector, respectively.

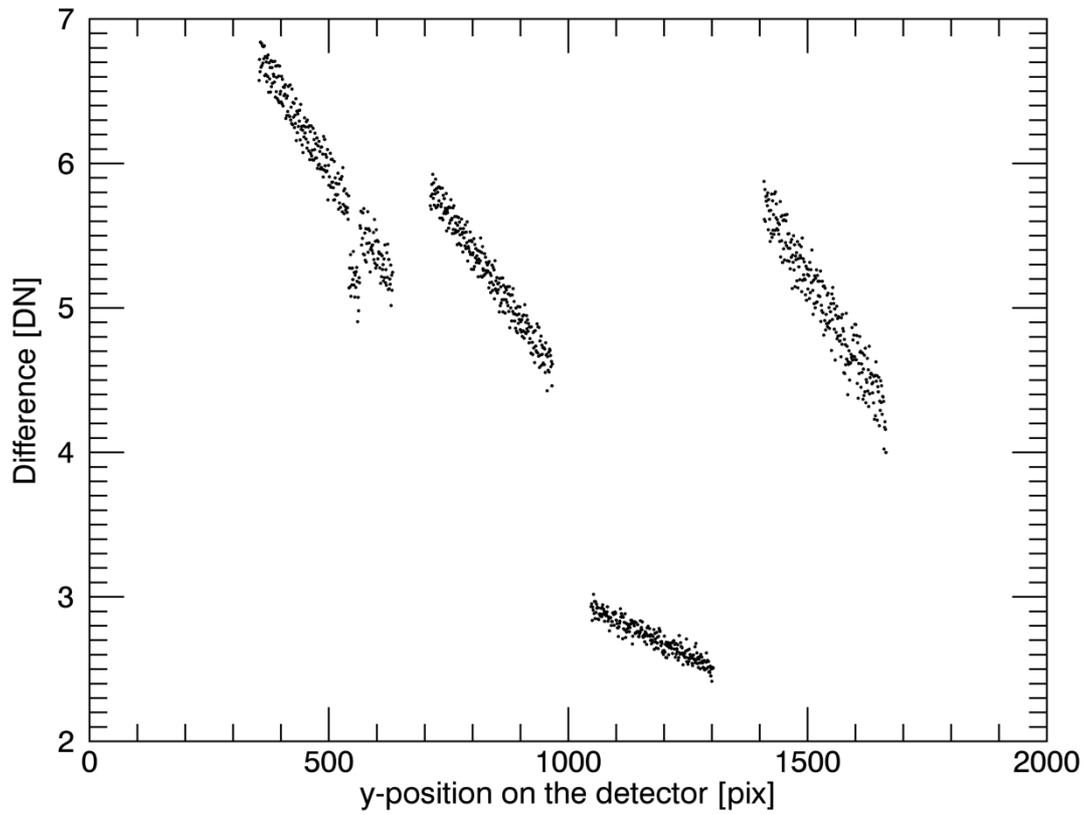

**Figure 5**: Difference (in DN) between the averaged y-profiles of the bias on the frame currently in use (20190308) and the newest product (20191220). The differences are small overall and show a slight linear gradient along the y-axis between the two frames with a maximum amplitude below 2DN within each individual window.

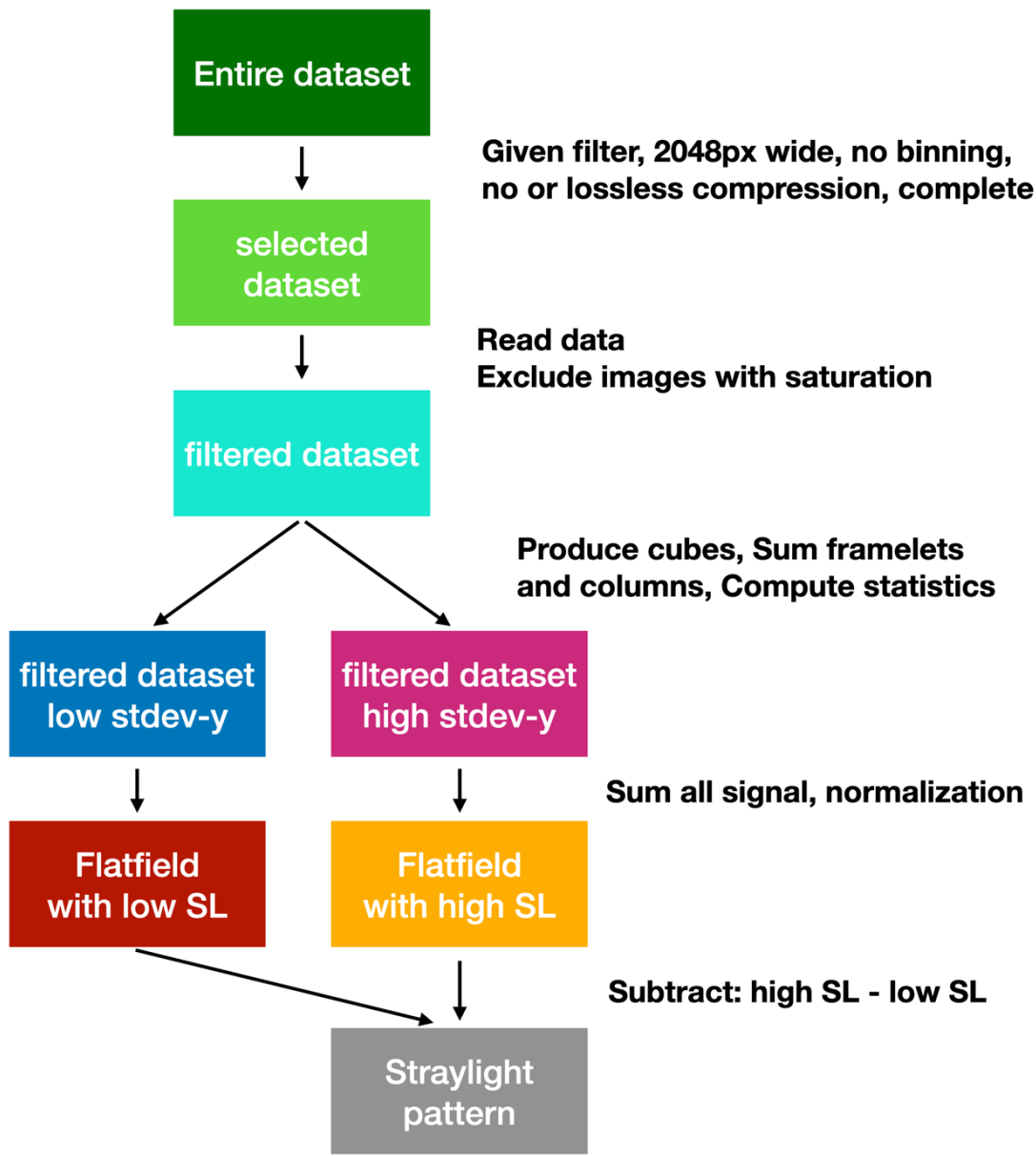

**Figure 6**: Flowchart of the pipeline used to generate the flatfield and the straylight pattern frames from a selection of images from the entire CaSSIS dataset.

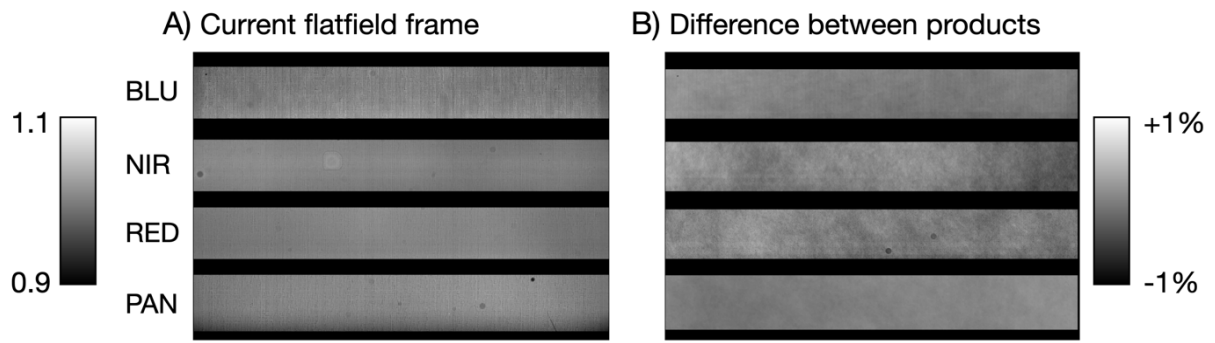

**Figure 7:** A) The flatfield frame currently in use in the CaSSIS photometric calibration pipeline (product 20190308) stretched linearly between 0.9 and 1.1. B) Difference between a flatfield frame generated from all data acquired before stp035 and a flatfield frame generated from all data acquired after the same stp using exactly the same method, as illustrated on Figure 4. The main difference between these two flatfield frames are the two shadows of dust grains toward the bottom and centre of the RED filter. The dust grain(s) responsible for the shadows has/have moved around the time of stp035. Beside this change, the only notable difference relates to the amount of straylight within the flatfield, visible in the form of darker bands toward the bottom of the RED and NIR filters.

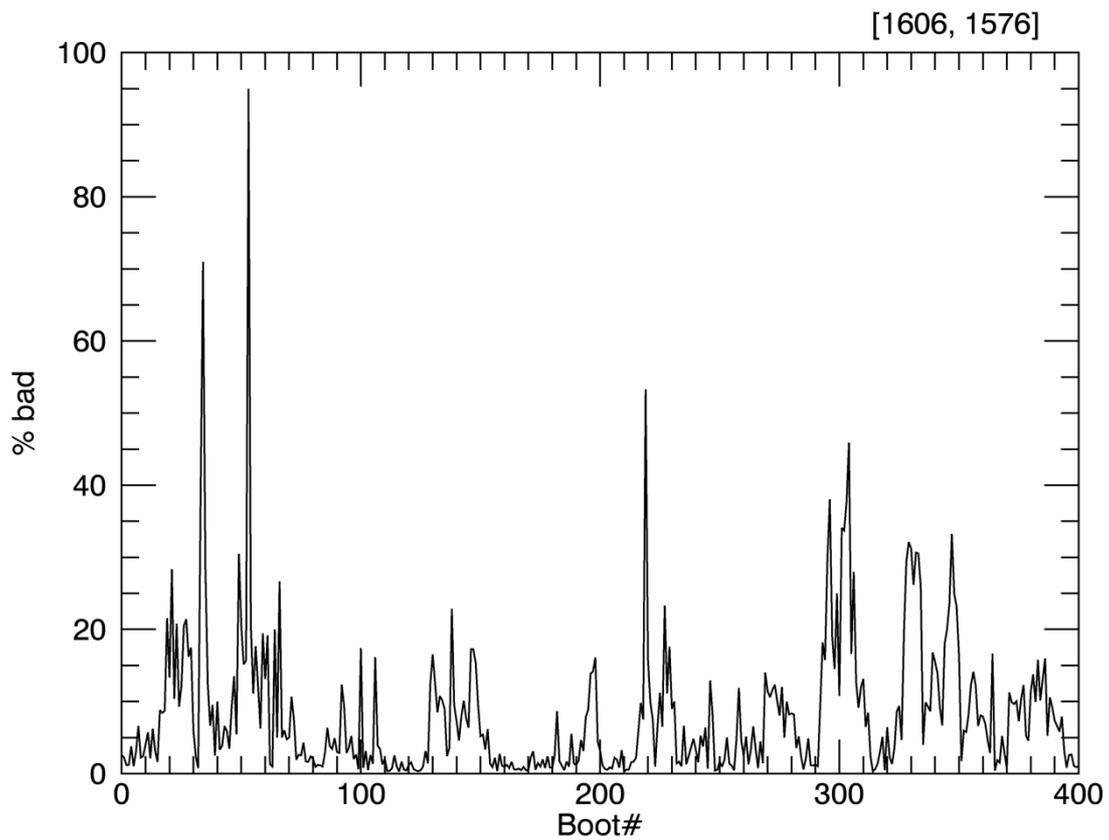

**Figure 8**: Temporal evolution of the failure rate for pixel [1606, 1576] located in the BLU window. See text for explanations of the way the failure rate is derived. The failure rate shows strong temporal variations throughout the mission, with periods where it stays close to 0 for several successive days of operations and short peaks where it exceeds 30% and reaches almost 100% failure rate in one occasion toward the beginning of the mission.

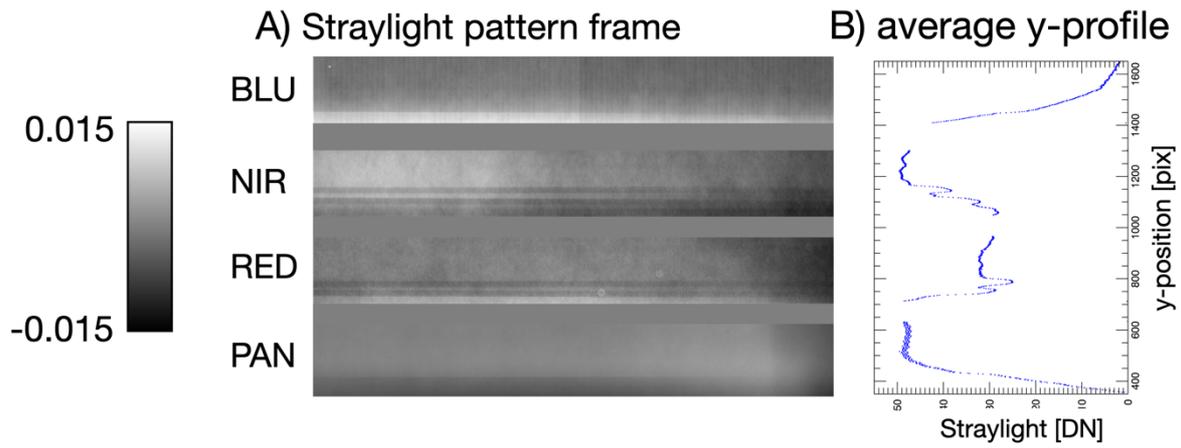

**Figure 9:** A) The straylight pattern frame currently in use in the CaSSIS photometric calibration pipeline (product 20190308) and produced using the method illustrated in Figure 4. B) Averaged y-profile of the straylight. The profiles in each filter have been stretched, offset and aligned based on the outputs of the straylight-pattern correction on the whole dataset.

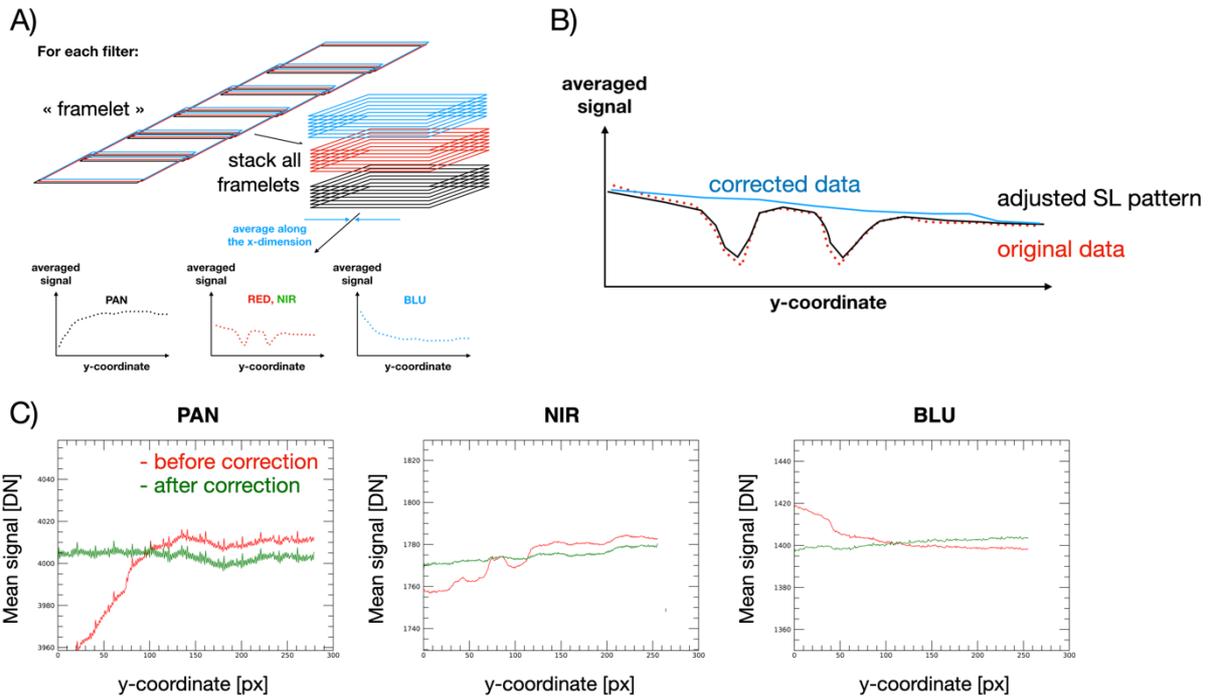

**Figure 10:** Sketch of the method used to remove the straylight pattern from the images. A) The correction algorithm is unidirectional, working with y-profile data averaged over all framelets for a given observation and filter and over the x dimension of the detector. B) The amplitude of the straylight pattern is then adjusted so that its removal leads to a y-profile as linear as possible, effectively removing as much as possible of the features from the straylight pattern. C) Example with actual data (observation MY34_002191_269_2 shown in Figure 9) showing the difference between the original (uncorrected) y-profile (in red) and the corrected profile (in green).

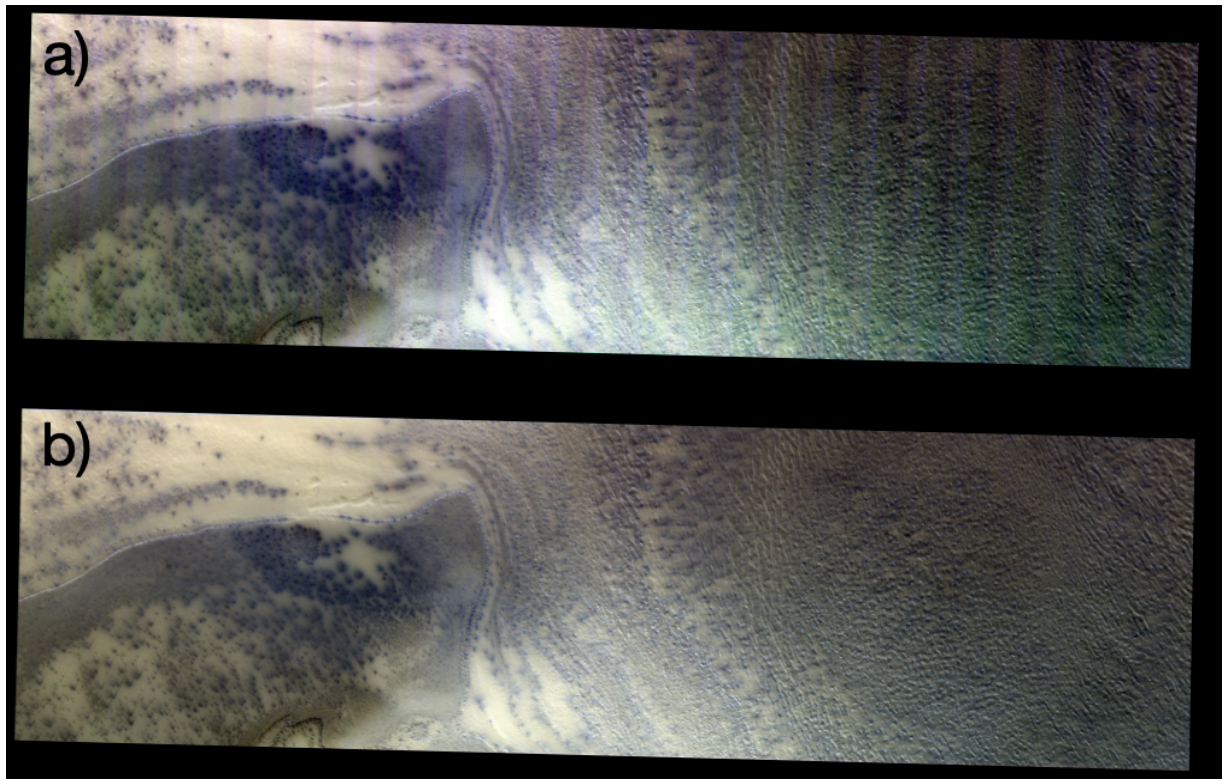

**Figure 11:** Example of result of the straylight-pattern removal (CaSSIS observation MY34_002191_269_2). a) Original calibrated image after bias subtraction and flatfield division. b) Same after removal of straylight and offsets. The straylight pattern removal uses the method illustrated in Figure 8C which shows the actual y-profiles plots for this particular observation. The additional correction steps of straylight gradients and framelet-to-framelet offsets are described in Section 3.6 and Figure 17.

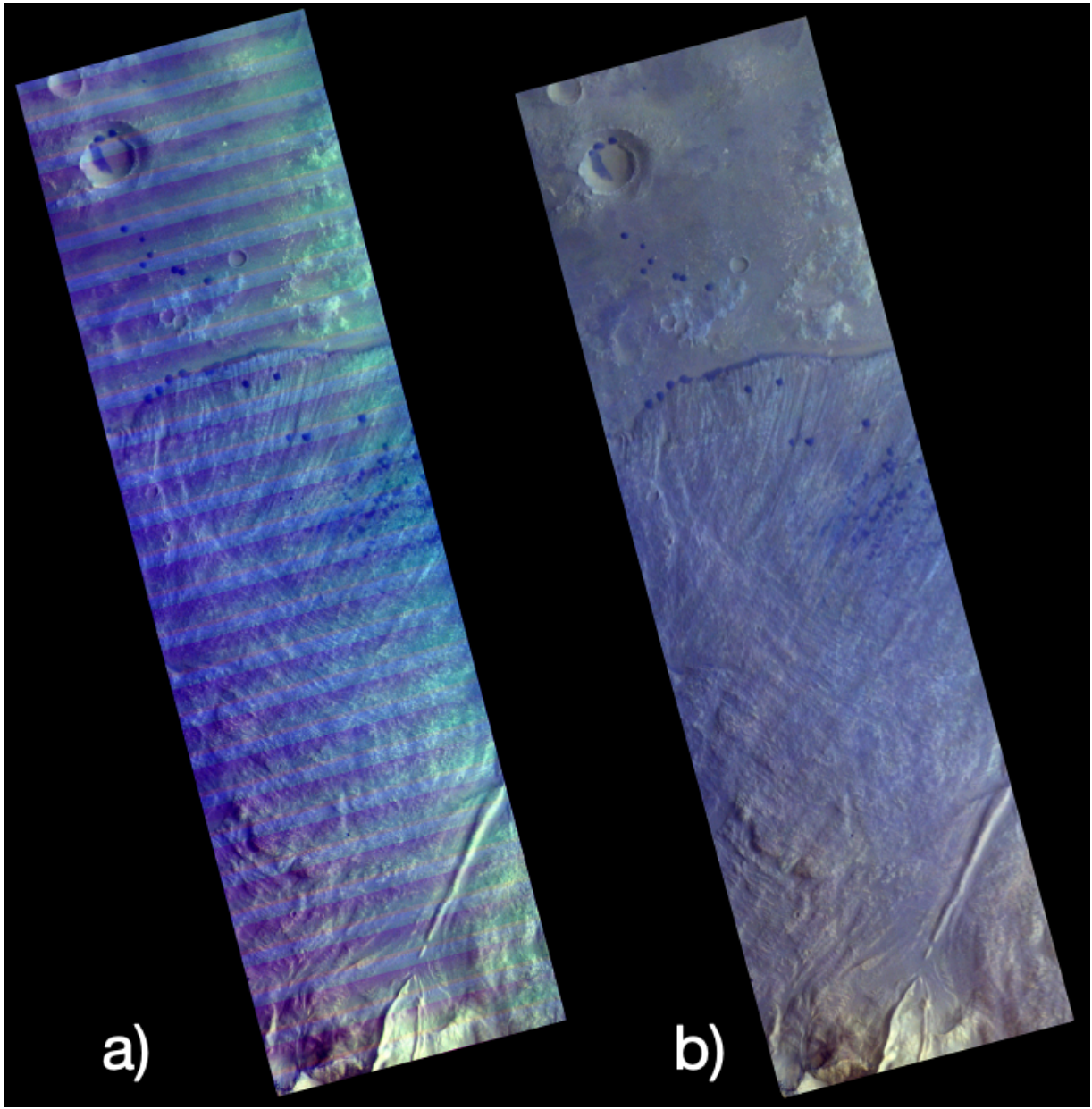

**Figure 12:** Example of results of the straylight correction (CaSSIS observation MY34_002084_210_2). a) Original calibrated image after bias subtraction and flatfield division. b) Same after removal of straylight and offsets. The straylight pattern removal uses the method illustrated in Figure 8C which shows the actual y-profiles plots for this particular observation. The additional correction steps of straylight gradients and framelet-to-framelet offsets are described in Section 3.6 and Figure 17.

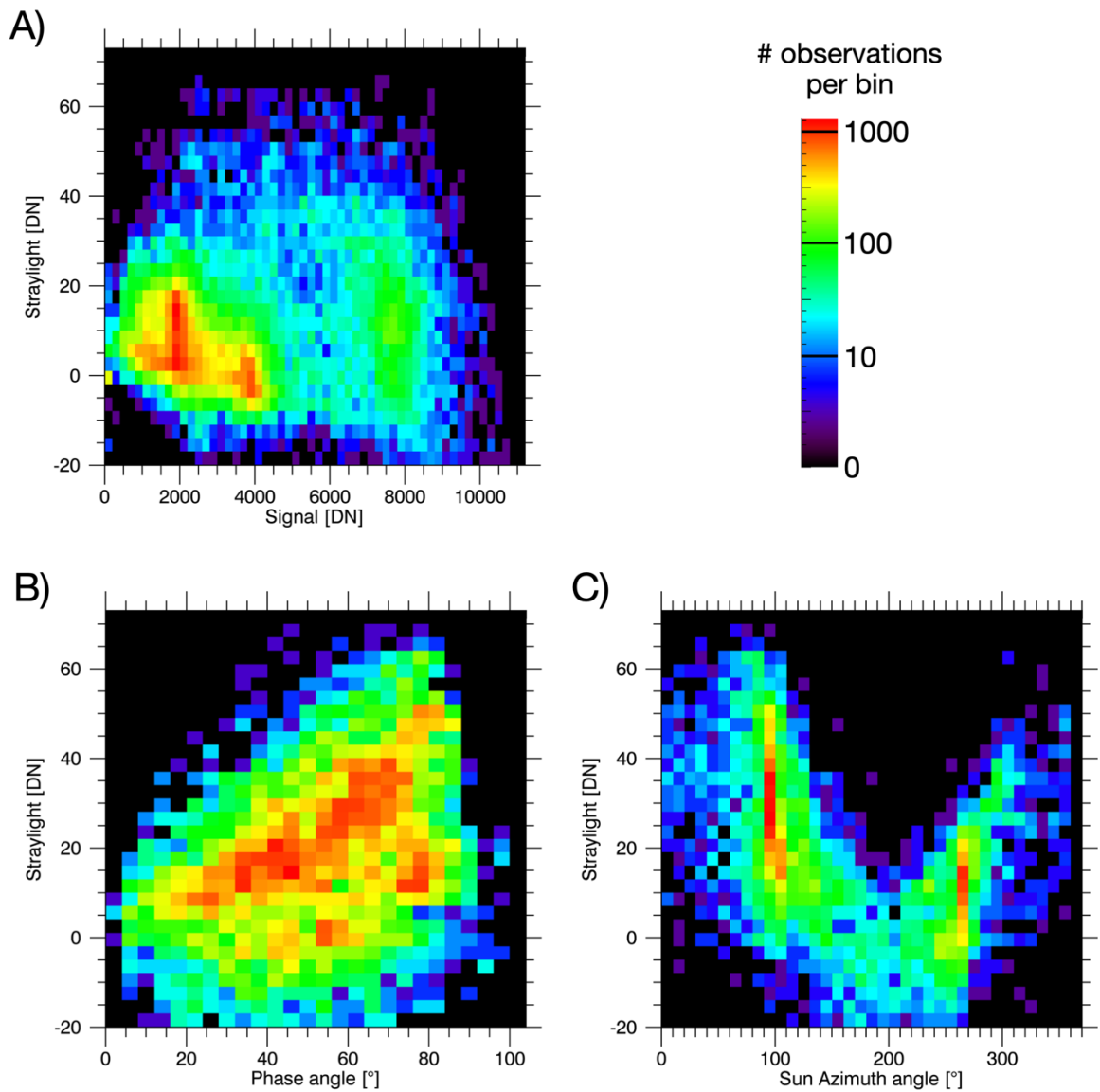

**Figure 13:** Results of the analysis of the straylight intensity across the entire CaSSIS dataset. The three scatterplots all present the amplitude of straylight (DN, y-axis) as a function of the actual signal (A), the Solar phase angle (B) and the Sun azimuth angle (C). The rainbow colour code indicates the density of data points in the scatterplot (warm colour: denser data points) as indicated on the colourbar. The scale is logarithmic.

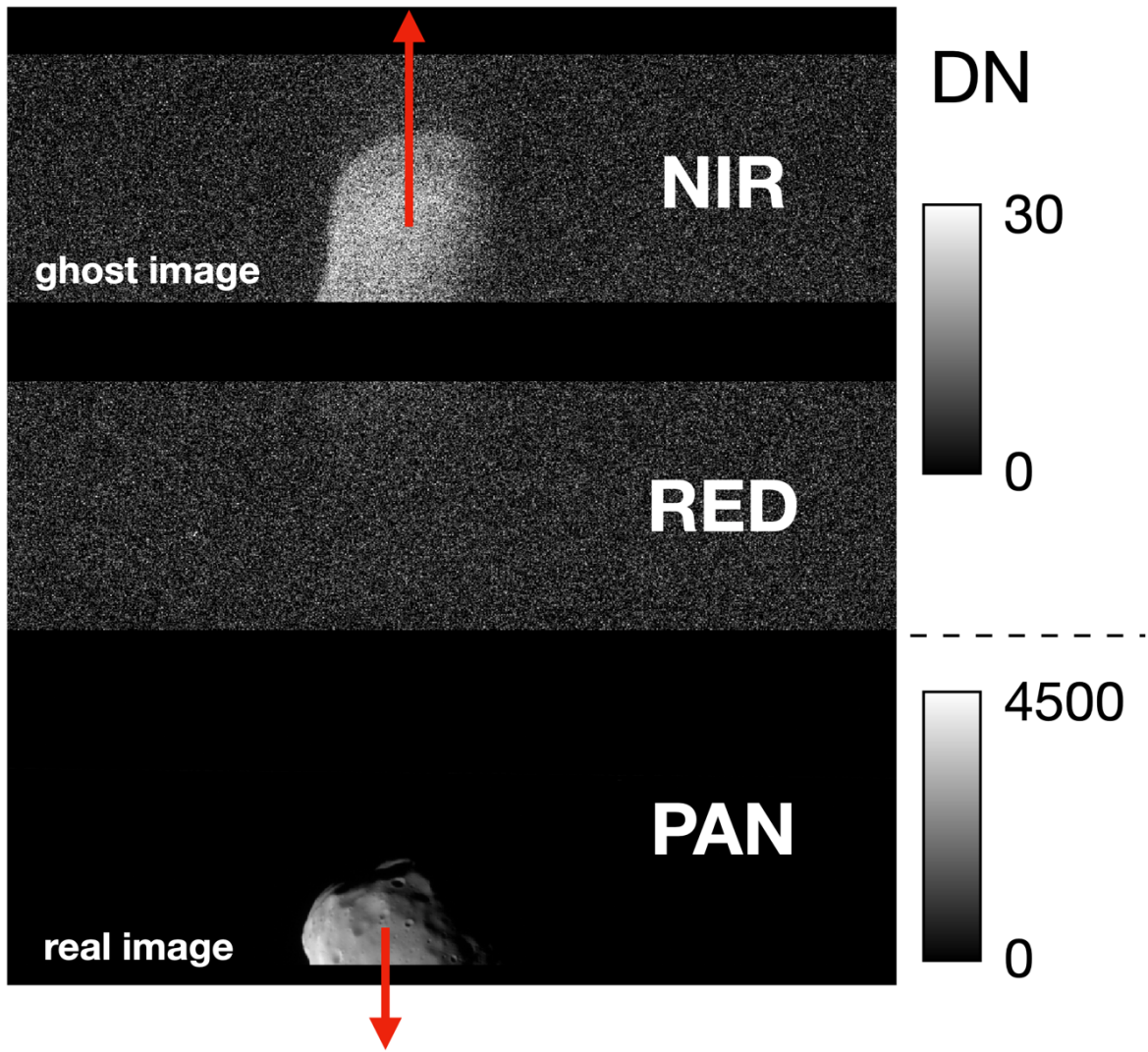

**Figure 14:** Ghost image of Phobos observed during the observation campaign from the initial capture orbit of TGO in November 2016 (Mars Capture Orbit #50). The actual image of Phobos appears at the bottom of the PAN filter while a relatively sharp and very faint ghost image of the out-of-field fraction of Phobos appears in the NIR filter. The images were bias subtracted and divided by the flatfield but not converted to I/F. The absolute intensity (in DN) of the ghost image in NIR reaches about 1% of the intensity of the actual image in PAN. Note the different stretches applied to the PAN filter (0 to 4500 DN) and to the NIR and RED filters (0 to 30 DN), as shown on the colourbars on the right. The red arrows indicate the opposite directions of motion of the real and the ghost image.

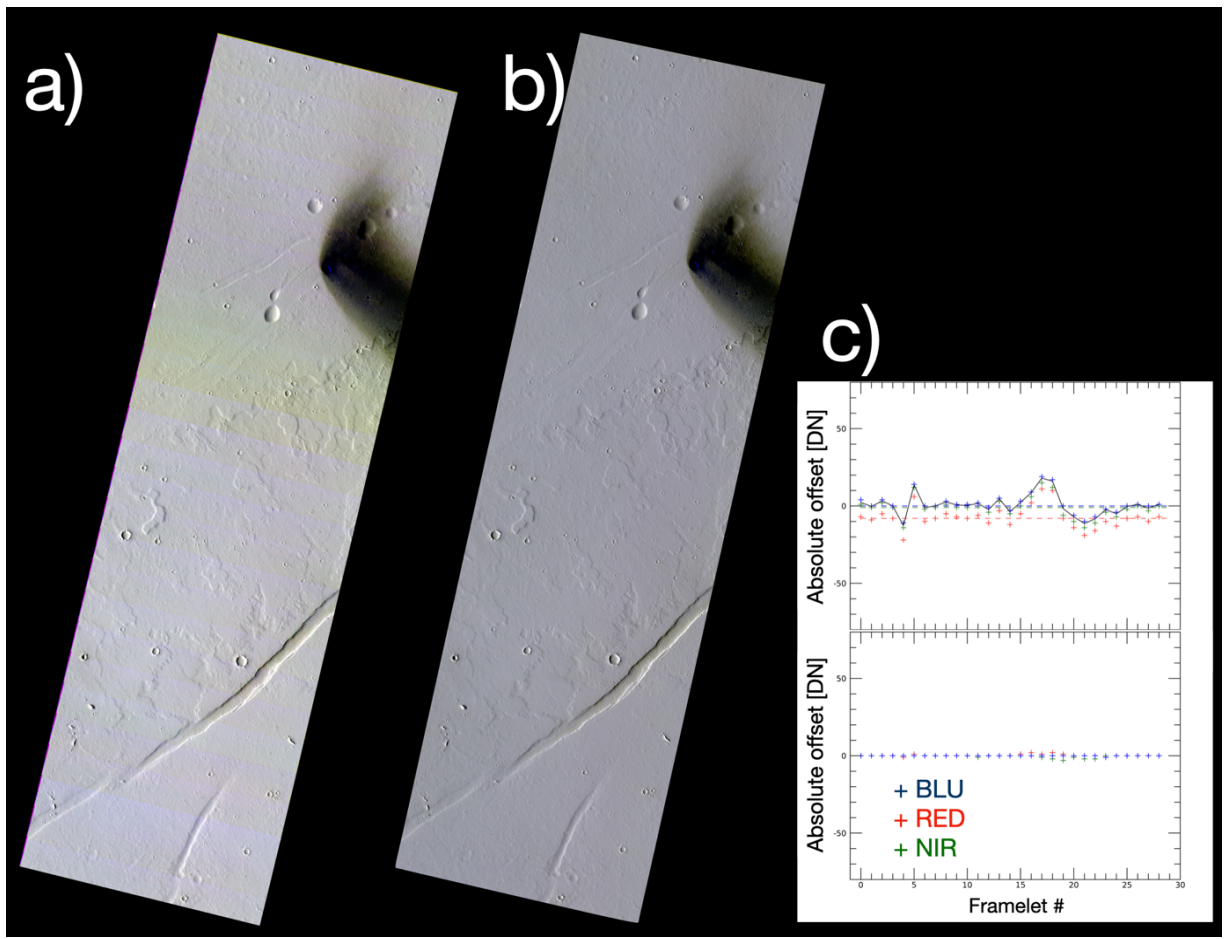

**Figure 15:** Example of correction of framelet-to-framelet offsets and straylight gradients (CaSSIS image MY34_002230_018_0). a) Colour composite assembled from the calibrated framelets before the correction of straylight and offsets. b) Same after correction of both effects. c) Plots showing the measured framelet-to-framelet offsets for all filters (top) before and (bottom) after corrections. The offsets are the difference in the median value over the overlapping region between adjacent framelets. They are caused by erratic variations in the bias level of the detector and therefore affect values in all filters similarly. A single subtractive correction from the average of values in all filters available is performed (see text for details).

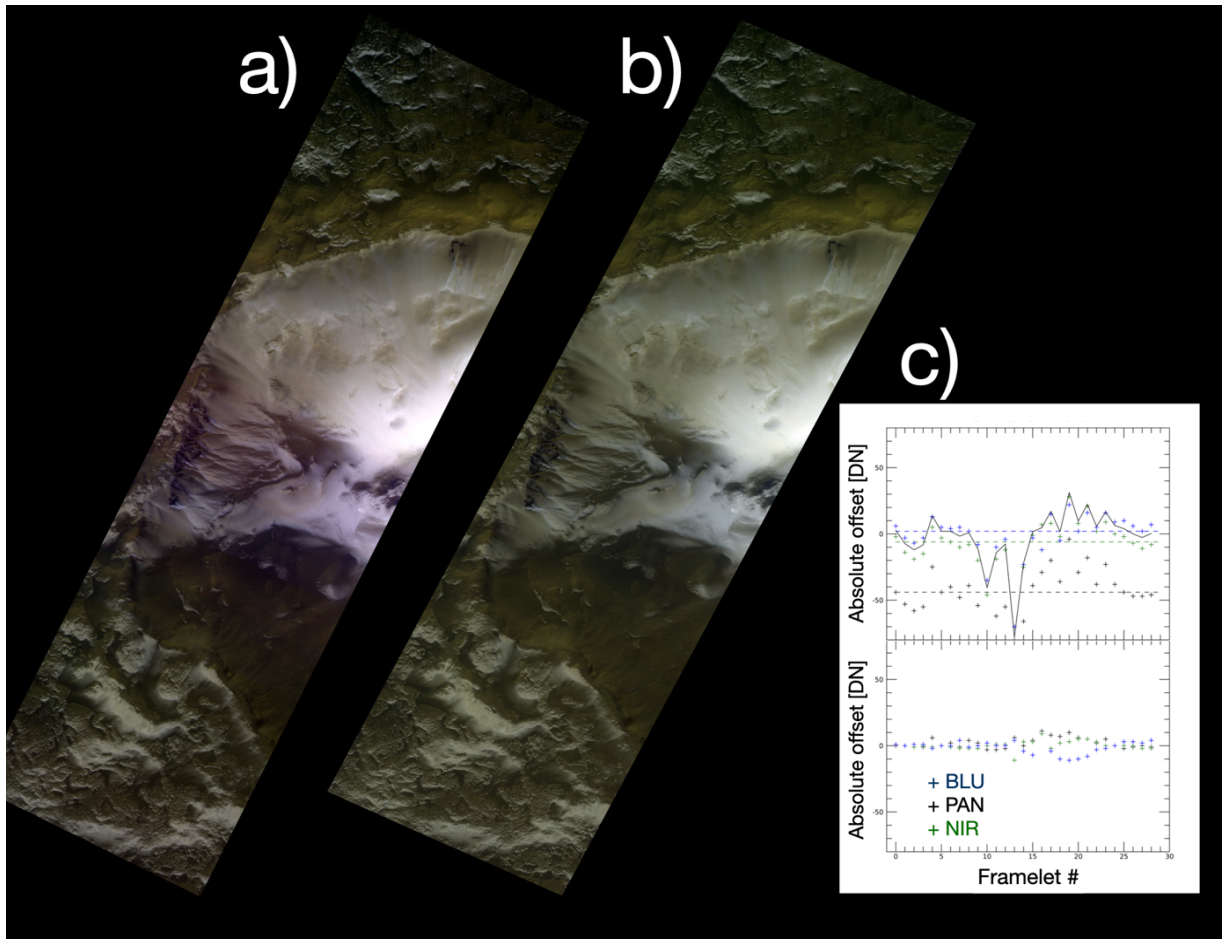

**Figure 16:** Example of correction of framelet-to-framelet offsets and straylight gradients (CaSSIS image MY34_002261_302_1). a) Colour composite assembled from the calibrated framelets before the correction of straylight and offsets. b) Same after correction of both effects. c) Plots showing the measured framelet-to-framelet offsets for all filters (top) before and (bottom) after corrections. The offsets are the difference in the median value over the overlapping region between adjacent framelets. They are caused by erratic variations in the bias level of the detector and therefore affect values in all filters similarly. A single subtractive correction from the average of values in all filters available is performed (see text for details).

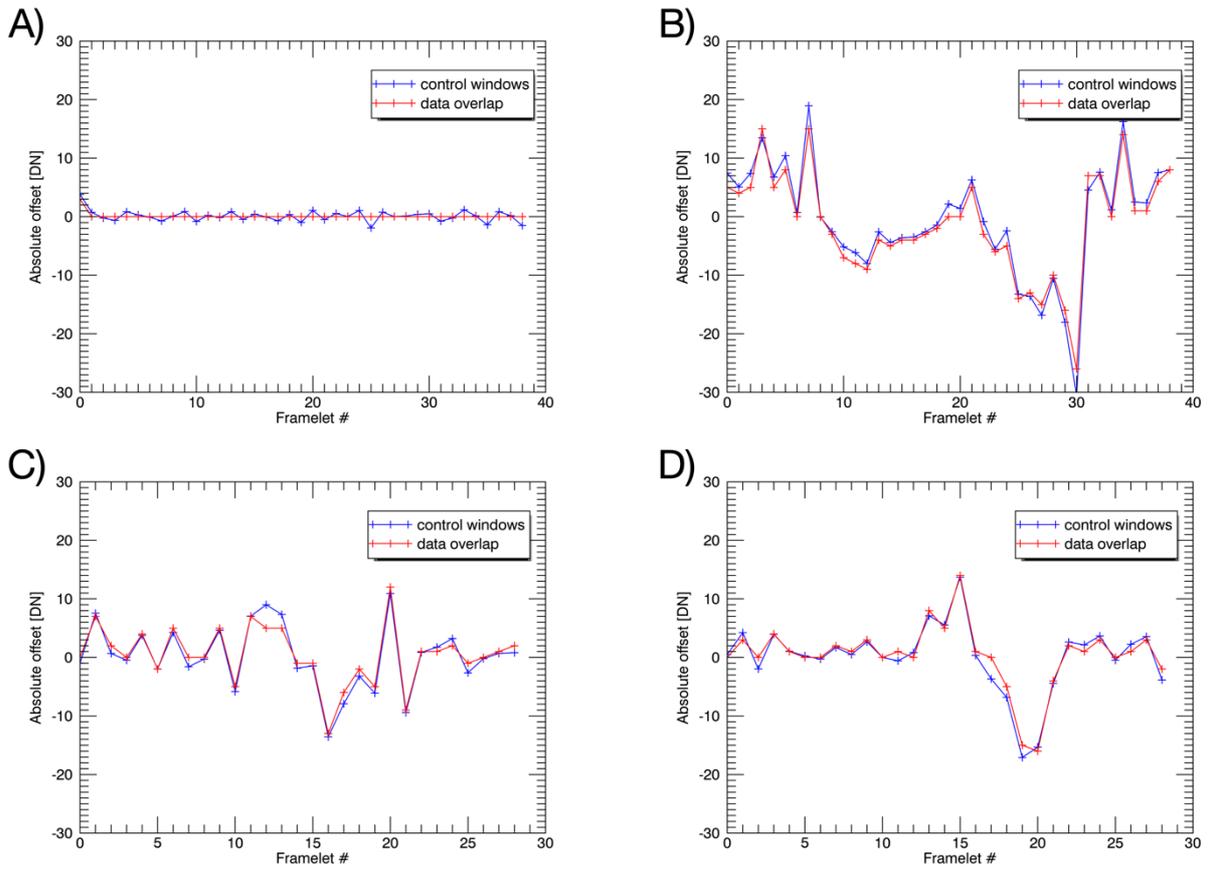

**Figure 17:** Comparison between two methods to determine the framelet-to-framelet offsets: the method employed here that utilizes the areas of overlap between adjacent framelets ("data overlap", see Figure 17) and the use of the two "control windows" placed under the mask that record changes of detector bias (Figure 1). Both methods show very consistent results, indicating that the observed framelet-to-framelet offsets are indeed caused by changes in detector bias and demonstrating that both methods can be used to characterize and correct these artefacts.

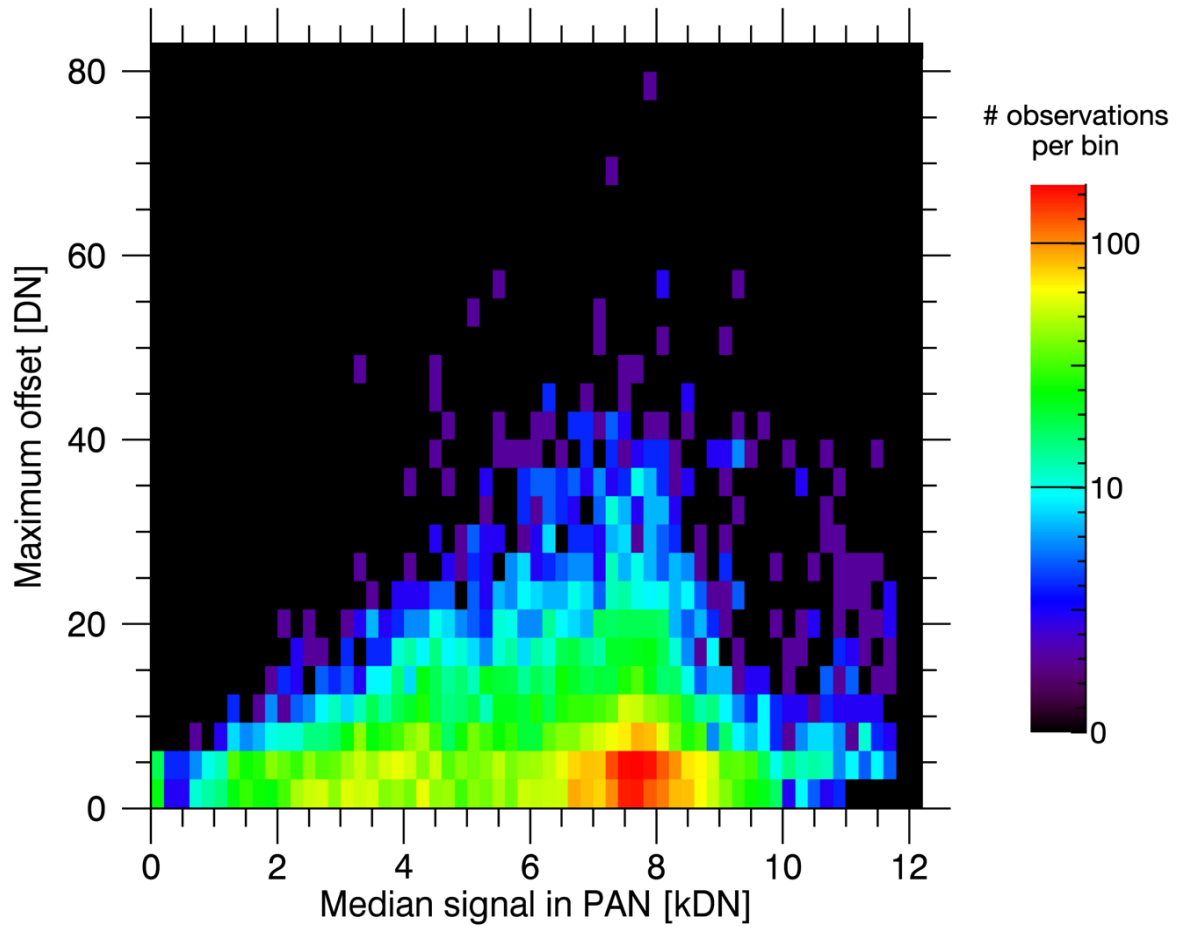

**Figure 18:** Statistics on the maximum framelet-to-framelet offsets found as a function of the median signal for the PAN filter. The scatterplot shows the amplitude of straylight (DN, y-axis) as a function of the median signal in PAN (DN, x-axis). The rainbow colour code indicates the density of data points in the scatterplot (warm colour: denser data points) as indicated on the colourbar. The scale is logarithmic.

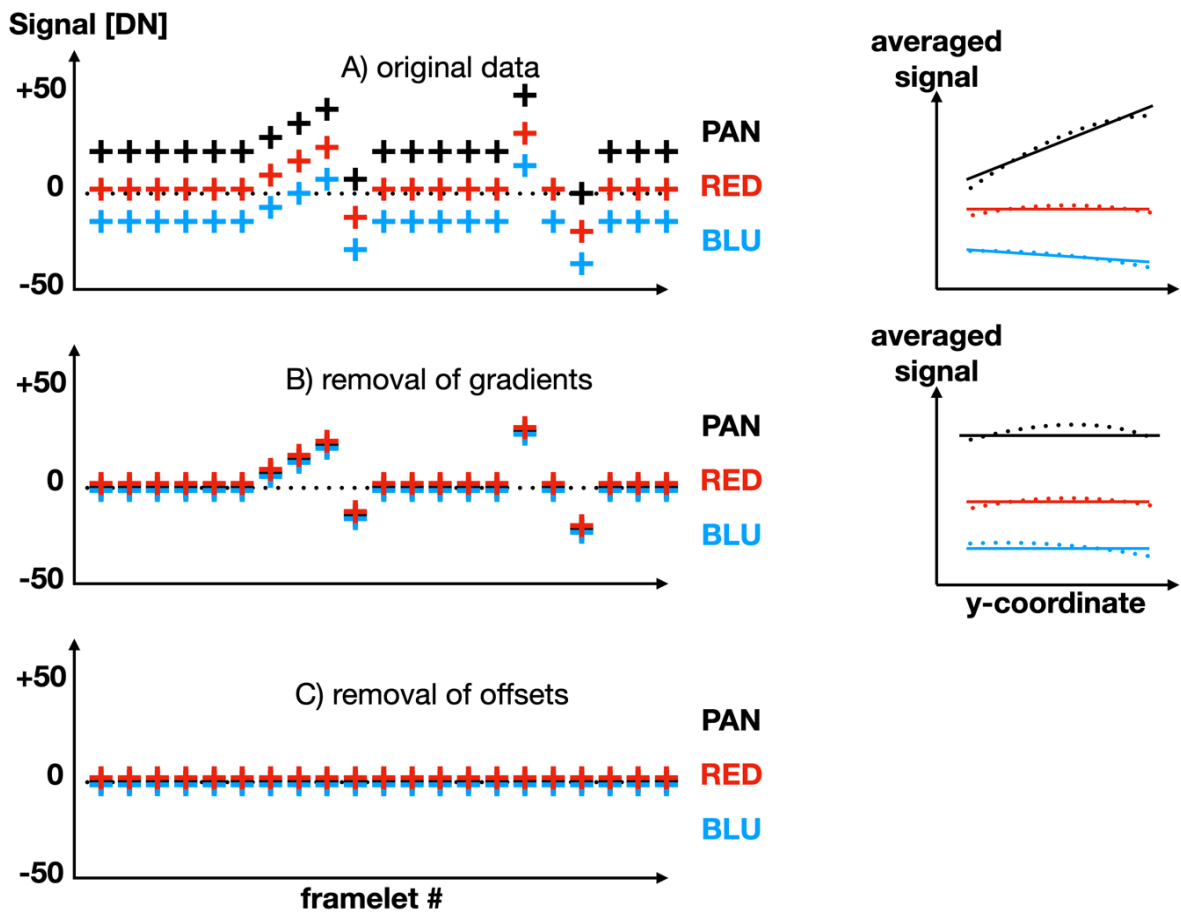

**Figure 19:** Sketch illustrating the principle of correction of the gradients and framelet-to-framelet offsets from the overlapping areas in the data. Left: plots showing the difference of signal (DN) over overlapping areas between adjacent framelets. Note that for each framelet #, the data through the different filters are acquired simultaneously but this corresponds to different areas of the surface. Right: y-profiles of the signal averaged along the x-dimension of the detector and all framelets.

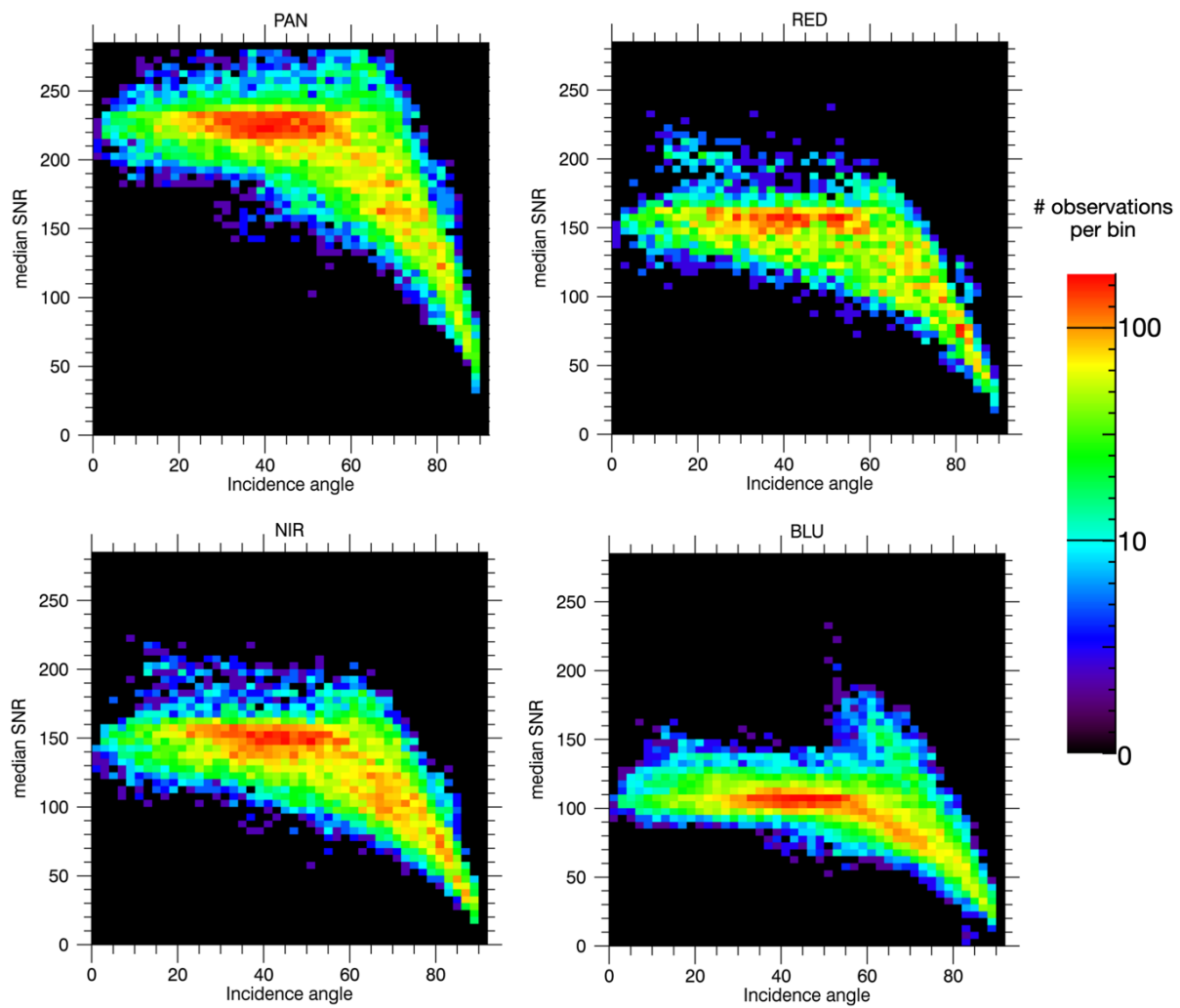

**Figure 20:** Density plots for the distribution of the median signal-to-noise ratio (SNR) of all CaSSIS images as a function of the incidence angle for the four different filters. The rainbow colour code indicates the density of data points in the scatterplot (warm colour: denser data points) as indicated on the colourbar. The scale is logarithmic. At incidence angle lower than about 40 to 60° depending on the albedo of the surface, the exposure time can be adjusted to obtain a median signal in PAN of 8000 DN which results in a SNR of 230 (taking into account both shot-noise and readout noise). In these conditions, the SNR is around 150 in RED and NIR and slightly above 100 in BLU. As incidence angle increases beyond ~60°, the exposure time becomes limited by the 1-pixel smearing time (1.5 ms) and the signal decreases with the cosine of the incidence angle, reducing the SNR. At very high incidence angle (>80°) the contribution of the scattering by atmospheric aerosols (clouds and dust) to the signal becomes very important and the signal diverges from cos i significantly.

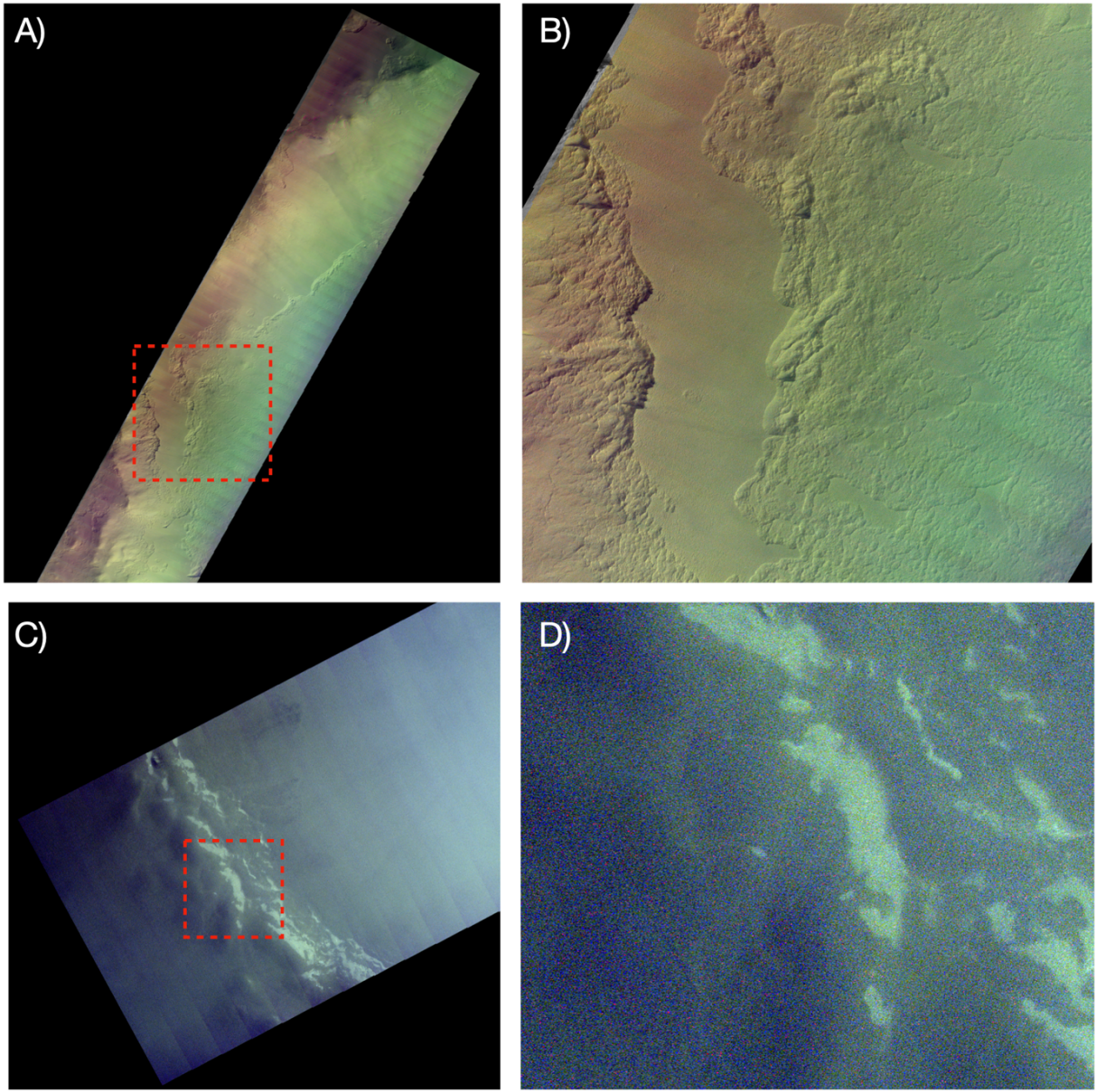

**Figure 21:** Examples of two CaSSIS images taken under low illumination conditions, which still show prominent calibration artefacts after correction. A) CaSSIS image MY35_011880_061_0 (NIR-PAN-BLU composite) of the western rim of Milankovic crater in the northern plains (212.1°E. 54.8°N). Uncorrected straylight results in a repetitive pattern seen mostly on both sides of the image. In addition, the left side of the image looks reddish while the right side looks bluish. This image was taken in very low light conditions (incidence angle: 87°). Zooming-in on the image (B) reveals that geologic features are actually recognizable but all colour information is lost into the noise. C) CaSSIS image MY34_001743_078_1_RPB (RED-PAN-BLU composite) of the southern rim of an unnamed crater in the northern plains (251.7°E, 69.9°N) taken at very early local solar time (04:17:31) with an incidence angle of exactly 90°. Zooming-in on this image (D) shows the high noise in the colour but still shows a contrast of colour between white ice and reddish dust.